\documentclass{ar2e}
\usepackage{graphicx}
\usepackage{rotating}
\usepackage{amsmath,amssymb}
\begin{document}
\input epsf.tex    

\input psfig.sty

\jname{Annu. Rev. Astron. Astrophys.}
\jyear{2011}
\jvol{49}
\ARinfo{1056-8700/97/0610-00}

\title{The First Galaxies}

\markboth{Bromm \& Yoshida}{First Galaxies}

\author{Volker Bromm$^1$ and Naoki Yoshida$^2$
\affiliation{$^1$ Department of Astronomy and Texas Cosmology Center, University of Texas, Austin, Texas 78712;
 email: vbromm@astro.as.utexas.edu\\
$^2$ Institute for the Physics and Mathematics of the Universe, University of Tokyo, Kashiwa, Chiba 277-8583, Japan;
 email: naoki.yoshida@ipmu.jp}}


\begin{abstract}
We review our current understanding of how the first galaxies formed at the
end of the cosmic dark ages, a few 100 million years after the Big Bang.
Modern large telescopes discovered galaxies at redshifts greater than seven,
whereas theoretical studies have just reached the degree of sophistication
necessary to make meaningful predictions. 
A crucial ingredient is the feedback exerted by the first generation of stars, 
through UV radiation, supernova blast waves, and chemical enrichment. 
The key goal is to derive the signature of the first galaxies to be 
observed with upcoming or planned next-generation facilities, 
such as the {\it James Webb Space Telescope} or 
{\it Atacama Large Millimeter Array}. 
From the observational side, ongoing deep-field searches for very 
high-redshift galaxies begin to provide
us with empirical constraints on the nature of the first galaxies.
\end{abstract}

\maketitle

\section{INTRODUCTION}

The first galaxies have captivated theorists and observers alike for more
than four decades. They were recognized as key drivers of early cosmic evolution
at the end of the cosmic dark ages, when the Universe was just a few 100 million
years old (e.g., Rees 1993; Barkana \& Loeb 2001). Within the standard 
$\Lambda$ cold dark matter ($\Lambda$CDM) cosmology, 
where structure forms hierarchically through mergers of smaller dark matter (DM)
halos into increasingly larger ones, the first galaxies were the basic
building blocks for galaxy formation (e.g., Blumenthal et al. 1984; 
Springel et al. 2005). The highly complex physics associated with galaxy assembly
and evolution still largely defies our understanding, but the first galaxies
may provide us with an ideal, simplified laboratory for study 
(e.g., Loeb 2010).

A crucial ingredient to any theory of how the first galaxies assembled, and how
they impacted subsequent cosmic history, is the feedback exerted by the
stars formed inside them or their smaller progenitor systems (e.g., Wise \& Abel
2008; Greif et al. 2010). Understanding the first galaxies is therefore
intricately linked to the formation of the first, so-called Population~III 
(Pop~III) stars (Bromm et al. 2009). The stellar feedback is usually divided
into radiative and supernova (SN) feedback (Ciardi \& Ferrara 2005).
The radiative effect consists of the build-up of H{\sc ii} regions around individual
massive Pop~III stars, thus initiating the extended process of cosmic
reionization (Sokasian et al. 2004; Barkana \& Loeb 2007). 
The SN feedback has a direct mechanical
aspect, where the blastwave triggered by the explosion imparts heat and 
momentum to the surrounding intergalactic medium (IGM). 
Supernovae also disperse heavy elements into the IGM, thereby affecting
the subsequent mode of star formation in the polluted gas. 
An early episode of enriching the primordial, pure H/He
Universe with metals is therefore another long-term legacy left behind
by the first stars and galaxies, together with reionization.

There is a further, observational, reason for the current flurry of
activity in understanding the first galaxies. Researchers wish to predict the
properties of the sources to be probed with upcoming or planned
next-generation facilities, such as the {\it James Webb Space Telescope
(JWST)}, the {\it Atacama Large Millimeter Array} (ALMA),
or extremely large telescopes to be constructed on the ground.
The main efforts in the latter category are the 
{\it Giant Magellan Telescope}, 
the {\it Thirty Meter Telescope}, 
and the {\it European Extremely Large Telescope},
which are pursued concurrently at the present time. In each case, we need
to work out the overall luminosities, spectral energy distributions or colors,
and the expected number densities of sources as a function of redshift.
Complementary to the direct detection approach are possible signatures
of the first galaxies in the redshifted 21-cm background radiation (Furlanetto 
et al. 2006; Morales \& Wyithe 2010). Here, 
a number of already operational or planned meter-wavelength radio telescopes,
among them the Low Frequency Array (LOFAR), will soon commence 
the search for the 21-cm signatures.
The effort of arriving at robust predictions for these facilities
greatly benefits from recent advances in supercomputer technology, where 
large- (tera and peta-) scale, massively parallel systems provide us with
unprecedented computational power to carry out ever more realistic
simulations in the cosmological context.

Most large galaxies today harbor supermassive black holes (SMBHs) in their centers
(e.g., Kormendy \& Richstone 1995; Ferrarese \& Ford 2005). An important question then is when and how galaxies
first acquired such central black holes. Related is the problem of understanding
the presence of $\sim 10^9 M_{\odot}$ SMBHs that are inferred to power the
luminous quasars discovered by the Sloan Digital Sky Survey (SDSS)
at redshifts $z \gtrsim 6$ (Fan, Carilli \& Keating 2006). 
A popular theoretical model assumes that such very massive black holes grew 
from smaller seeds, present already in the smaller progenitor
systems that merged into the massive SDSS quasar hosts (Li et al. 2007). 
The efficieny of growing a black hole via accretion of surrounding gas over the available
time of several hundred million years, however, may have been quite limited. 
A possible way out is to begin the SMBH assembly process already with more massive seeds.
The first galaxies have indeed been suggested as viable formation sites
for such $\sim 10^6 M_{\odot}$ seed black holes (see Section~5).
Regardless of their exact properties and origin, such massive black holes 
would likely have influenced the structure and evolution
of the first galaxies (e.g., Cattaneo et al. 2009).

The nature of the stellar populations in the first galaxies is crucial
for the observational quest. According to some theories,
the majority of the first galaxies already contained low-mass,
Population~II (Pop~II), stars, and perhaps stellar clusters in general.
This expectation is based on the theory of a `critical metallicity',
$Z_{\rm crit}\sim 10^{-6}-10^{-4} Z_{\odot}$, above which the mode of
star formation is thought to change from top-heavy to normal, bottom-heavy 
(e.g., Bromm et al. 2001; Schneider et al. 2002). 
Due to the pre-enrichment from Pop~III stars in the galaxy's progenitor systems, 
the so-called minihalos (Haiman, Thoul \& Loeb 1996; Tegmark et al. 1997; 
Yoshida et al. 2003), 
the first galaxies were likely already supercritical, thus experiencing Pop~II 
star formation. 
The minihalos, consisting of DM halos with total mass $\sim 10^{6} M_{\odot}$ and 
collapsing at $z\sim 20-30$, are the formation sites for the first (Pop~III)
stars. Cooling inside of them relies on a trace amount ($\sim 10^{-3}$ by number)
of molecular hydrogen. These halos have shallow potential wells, so that
they are highly susceptible to negative feedback effects from Pop~III stars.
A subset of the first Pop~II star, those with subsolar masses, 
will survive to the present, and can thus be probed as fossils of the dark ages 
in our immediate cosmic neighborhood. This approach, often termed stellar
archaeology (e.g., Beers \& Christlieb 2005; Frebel 2010),
provides constraints on the SN yields of the first stars, as well as
on the environment for star formation inside the first galaxies. A similar
strategy has recently become feasible, where the stellar content and
structural properties of low-mass dwarf galaxies in the Local Group
are interpreted under the assumption that they are descendants of
the first galaxies (e.g., Tolstoy, Hill \& Tosi 2009;
Ricotti 2010). Finally, these early galaxies are also discussed as
formation sites for the oldest globular clusters (Bromm \& Clarke 2002;
Kravtsov \& Gnedin 2005; Moore et al. 2006; Brodie \& Strader 2006; 
Boley et al. 2009; Cooper et al. 2010).

The plan for this review is as follows. We begin by considering the
seemingly straightforward question: What is the definition of a first
galaxy? It turns out that there is no universally accepted definition, as
is the case for what we mean by the formation of the first stars. Theorists
and observers employ different concepts, and often do not agree even among
themselves. We will try to clarify the situation (Section 2). We then turn
to a survey of what is known from existing observations that push the
envelope and begin to reach very high redshifts (Section 3).
This is followed by a more extended discussion of the lessons gleaned from
recent simulations, many of them studying the assembly of the first galaxies
with considerable physical sophistication and within a realistic
cosmological context (Section 4). Due to its importance, we devote a separate
section to the early co-evolution of massive black holes and stellar systems,
although our knowledge here also relies mostly on theory and numerical 
simulations (Section 5).
The following two sections discuss the observational signature of the
first galaxies, with a special focus on the {\it JWST}, but also
addressing the stellar archaeology approach, as well as more indirect
clues from the cumulative impact of the first galaxies on reionization,
21cm radiation, and the cosmic infrared background (Sections 6 -- 7).
We conclude with a brief outlook into the exciting decade ahead.

At the end of this introduction, we would like to point the reader to
a few other reviews that cover material related to this subject here.
For a general overview of the end of the cosmic dark ages, see the
extensive review by Barkana \& Loeb (2001), the more succinct one by
Bromm et al. (2009), and the monographs by Stiavelli (2009) and Loeb (2010). 
Feedback processes are discussed in detail by Ciardi \& Ferrara (2005), whereas
the physics and observational picture of reionization are treated by
Fan et al. (2006), Barkana \& Loeb (2007), and Meiksin (2009).
The formation of the first stars was reviewed by Bromm \& Larson (2004)
and Glover (2005). It is instructive to consider the huge lore of
knowledge that we have on present-day star formation, when extrapolating
to the primordial case. Comprehensive resources are the reviews by
McKee \& Ostriker (2007) and Zinnecker \& Yorke (2007).
The field of stellar archaeology has been summarized by Freeman
\& Bland-Hawthorn (2002), Beers \& Christlieb (2005), 
Tolstoy, Hill \& Tosi (2009), Frebel (2010), and Ricotti (2010).
Finally, Mo, van den Bosch \& White (2010) have written an excellent textbook
that summarizes all aspects of galaxy formation and evolution in the
proper cosmological context (also see Benson 2010).

\section{WHAT IS A FIRST GALAXY?}

There is currently no universally agreed upon definition of what we mean by ``first galaxy''.
Observers and theorists operate with different working hypotheses, and those hypotheses
have changed with our evolving understanding. We here summarize the most common attempts
to define a primordial galaxy. Intriguingly, to properly pose the question (``What is a
first galaxy?''), we already need to know the answer to it. It is thus likely, that we
will witness a continuing iterative process, but it is also evident that devising a
proper definition must be part of the discovery process.

\subsection{Theoretical Perspective}

On the theory side, the discussion typically begins with an enumeration 
of defining properties. 
What are the ingredients required for a first galaxy ? For a galaxy in general, 
the presence of a confining dark matter
halo hosting a long-lived stellar system seems inevitable. Often, there is gas present as 
well, but there are galaxies without any apparent gas. In addition, we may stipulate that
the potential well of the DM halo is sufficiently deep to retain gas that was 
heated to temperatures in excess of $\sim 10^4$\,K as a result of photo-ionization
by stellar radiation (Mesinger \& Dijkstra 2008; Mesinger, Bryan \& Haiman 2009).
More stringently, we may also want to demand that the halo can retain gas heated
and accelerated through SN explosions. Finally, we may ask whether the system is
able to support a multi-phase interstellar medium, 
which in turn could sustain a stable mode of self-regulated star formation.

The theorists' debate now centers on identifying the smallest, lowest-mass, DM halos that
fulfill the criteria listed above. 
According to the tentative list of criteria for what constitutes a galaxy, 
DM halos that are unable to form stars and therefore remain dark would not 
be galaxies. This would in particular apply to the first DM halos to collapse, 
or virialize, at $z\sim 100$. Their mass scale strongly depends on the nature 
of the CDM particle (Diemand, Moore \& Stadel 2005). For DM consisting of weakly 
interacting massive particles, as predicted by the theory of supersymmetry, 
the first DM halos comprise a mass of $\sim 10^{-6} M_{\odot}$, roughly 
the mass of Earth. For axion DM, on the other hand, the smallest halos
would only contain $\sim 10^{-13} M_{\odot}$.
Returning to the discussion of those DM halos that are able to form 
stars, one class of models proposes minihalos, defined in Section~1, as hosts for
the first galaxies (Ricotti, Gnedin, \& Shull 2002a, 2002b, 2008). 
In this case, the halos that host the
formation of the first (Pop~III) stars would coincide with the first galaxies. This
{\it Ansatz}, however, makes the implicit assumption that the initial mass function (IMF)
of the first stars was not very different from the locally observed one, where the
distribution peaks at low masses around $< 1 M_{\odot}$. 
Negative feedback effects from them, in terms of star-formation efficiency, 
would not be so severe for a subset of minihalos, 
such that they could sustain star formation and effectively self-enrich. 

Assuming that primordial stars were predominantly massive,
as is suggested by most current theoretical models and simulations (Omukai \& Palla 2003;
Bromm et al. 2009), leads to a very different picture, though. After the first stars are 
formed inside a minihalo, vigorous negative feedback effects would effectively shut off the potential
for subsequent star formation. For once, the heating due to photo-ionization drives a
pressure wave that greatly suppresses the gas density inside of minihalos
(Kitayama et al. 2004; Whalen, Abel \& Norman 2004; Alvarez, Bromm \& Shapiro 2006). 
If, in addition, energetic SNe occurred, the minihalo would be virtually 
devoid of any gas, leaving behind a sterile system as far as
star formation is concerned (Bromm, Yoshida \& Hernquist 2003; Greif et al. 2007). 
More massive systems that are able to re-assemble
the high-entropy material affected by Pop~III stars inside minihalos might
therefore be needed. There are, however, studies of the SN feedback in
primordial minihalos that suggest a different conclusion (Whalen et al. 2008).
If the bulk of the minihalo were to remain substantially neutral, thus
not triggering such dramatic outflows and the corresponding density suppression,
the SN remnant would be highly radiative and 
largely confined to the minihalo, thus effectively
self-enriching them. The condition of near-neutrality would be satisfied
in more massive ($\sim 10^7 M_{\odot}$) minihalos (Kitayama \& Yoshida 2005), 
combined with not too massive Pop~III progenitor stars. 
It is an open question whether these
conditions are ever met in a realistic cosmological setting, where
Pop~III star formation first occurred in lower-mass systems.

To gauge how susceptible a given halo will be to negative stellar feedback, it
is useful to introduce its virial temperature

\begin{equation}
T_{\rm vir}=\frac{\mu m_{\rm H} V_c^2}{2 k_{\rm B}}
\simeq 10^4\ \left(\frac{\mu}{0.6}\right)
\left(\frac{M}{10^8 M_{\odot} }\right)^{2/3}
\left[\frac{\Delta_c} {18\pi^2}\right]^{1/3}
\left(\frac{1+z}{10}\right)\ {\rm K},
\end{equation}

\noindent
where $V_c$ is the circular velocity, $\mu$ is the mean molecular weight,
and $\Delta_c$ gives the density contrast established
through virialization as a function of redshift (Bryan \& Norman 1998).
Closely related is the gravitational binding energy of the halo

\begin{equation}
E_b= \frac{1}{2}\frac{GM^2}{r_{\rm vir}} \simeq 5\times 10^{53}
\left(\frac{M}{10^8 M_{\odot} }\right)^{5/3}
\left[
\frac{\Delta_c} {18\pi^2}\right]^{1/3} \left(\frac{1+z}{10}\right)
 {\rm erg,}
\end{equation}

\noindent
where $r_{\rm vir}$ is the virial radius of the halo. In evaluating these
expressions, we have assumed cosmological parameters as recently
determined by the {\it Wilkinson Microwave Anisotropy Probe (WMAP)}
(Komatsu et al. 2009), and that $z\gg 1$.

\begin{figure}[ht]
\begin{center}
\includegraphics[angle=90,width = 5 in] {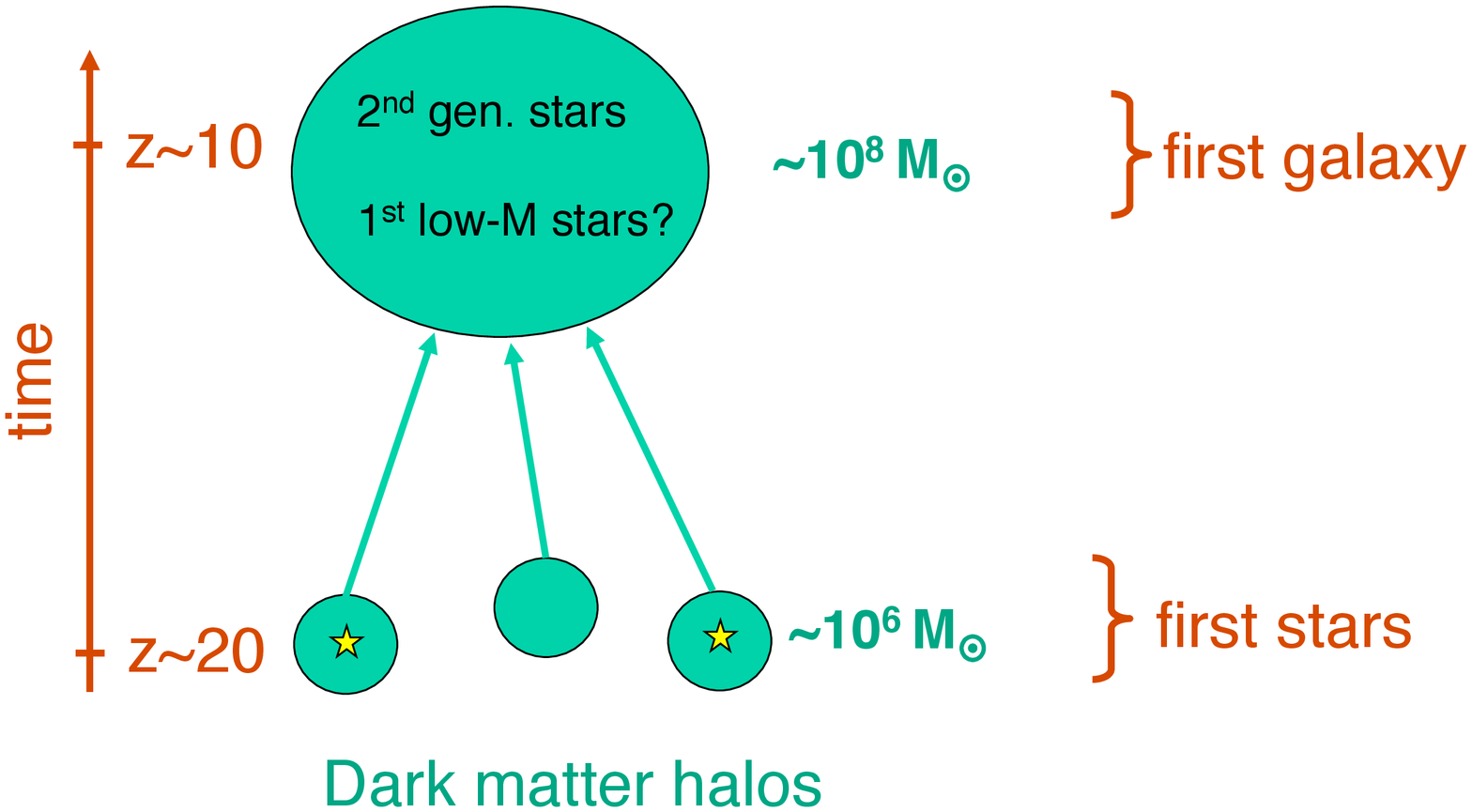}
\caption{Assembly of the first galaxy. We here illustrate the
scenario where the first galaxies reside in atomic cooling halos.
These comprise total masses of $\sim 10^8 M_{\odot}$ and typically
collapse at $z\sim 10$. 
Note that within the scenario illustrated in this figure, minihalos 
are not considered galaxies, because of the strong negative feedback 
from the Pop III stars that form inside of them. This feedback will 
effectively destroy the minihalos such that neither gas nor low-mass 
stars will remain in them. 
Their assembly is affected by the feedback
from the first (Pop~III) stars that had formed earlier in the
minihalo progenitor systems. Within this model, atomic cooling halos
hosted the second generation of stars, including the first low-mass
(Pop~II) stars that could have survived to the present day.}
\label{ASSEMBLY}
\end{center}
\end{figure}

Another series of recent simulations has suggested that DM halos containing a mass 
of $\sim 10^8 M_{\odot}$ and collapsing at $z\sim 10$ were the hosts for the first 
{\it bona fide} galaxies (Wise \& Abel 2007, 2008; Greif et al. 2008, 2010). These dwarf 
systems can indeed re-virialize the gas that was affected by previous star formation 
in minihalos (see {\bf Figure~\ref{ASSEMBLY}}). They are special in that their associated 
virial temperature exceeds 
the threshold, $\sim 10^4~\rm{K}$, for cooling due to atomic hydrogen (Oh \& Haiman 2002). 
These so-called `atomic-cooling halos' did not rely on the presence of molecular 
hydrogen to enable cooling of the primordial gas. In addition, their potential 
wells were sufficiently deep to retain photoheated gas, in contrast to the shallow 
potential wells of minihalos (Dijkstra et al. 2004). Our tentative conclusion is 
that atomic cooling halos thus seem to fulfill the requirements
for a first galaxy, but important open questions remain that need to be addressed 
with future simulations (see Section~4). 

A related issue is to identify the conditions
that enable the formation of the first disk galaxies (Pawlik, Milosavljevic \& Bromm 2011), 
or of central
supermasive black holes (see Section~5). However, such disks and central black holes may well
have emerged only at a later stage of hierarchical structure formation, {\it after} the first
galaxies had already formed. In this regard, they would not be necessary ingredients 
for a first galaxy, although they may well have been prevalent at the highest redshifts.

\subsection{Observational Perspective}

From the observational side, there are two main operational definitions employed.
One may simply equate ``first galaxy'' with the highest redshift galaxies observable
at a time, given its technology is pushed to the very limit. Currently, with a
combination of {\it Hubble Space Telescope} (HST) photometry and ground-based 
8-10m class spectroscopy, this
allows us to see galaxies at $z>7$, with a record of $z\simeq 8.6$ 
(Iye et al. 2006; Bouwens et al. 2010a; Lehnert et al. 2010),
or possibly even of $z \sim 10$ (Bouwens et al. 2011).
Evidently, this is a moving target, and such a temporary definition
makes it hard to provide a focus for theoretical studies.
In general, a number of galaxies at different evolutionary stages will
be present concurrently at a given redshift. 
Thus it would clearly be preferable if a definition involved some unambiguous criteria, 
based on the underlying physics.

A more precise definition is to search for galaxies
with zero metallicity, or one that hosts predominantly Pop~III stars. This
popular definition of a first galaxy may however be misleading, and may render any
attempts to find first galaxies futile from the very outset. 
This is because most first galaxies could be already metal-enriched
by SNe triggered by the first stars.
Recent simulations have indicated that heavy element production and dispersal was 
very rapid, leading to a bedrock of pre-galactic enrichment after only a few Pop~III 
stars had formed (see Section~4). Indeed, some models predict that the first 
galaxies predominantly already hosted Pop~II stars (Greif et al. 2010; Maio et al. 2011).
In summary, we will employ the following tentative definition of ``first
galaxy'' in this review: a galaxy comprised of the very first system of
stars to be gravitationally bound in a dark matter halo, regardless
of whether the stars are Pop~III or Pop~II.

In concluding this section, we would like to briefly comment on the concept
of a ``protogalaxy'', which is now largely only of historical interest. The idea
was that a mature galaxy like our Milky Way (MW) more or less evolved in a monolithic
fashion (Eggen, Lynden-Bell \& Sandage 1962; hereafter ELS), and not in the hierarchical,
bottom-up way that is now widely favored within the standard $\Lambda$CDM model. 
One could then go back in time, making predictions for the luminosity and color of
such systems during their initial, monolithic, collapse at high $z$ 
(Partridge \& Peebles 1967). First galaxy then referred to this initial collapse phase.
In many ways, defining, and understanding, the first galaxies 
in a hierarchical context is more difficult than
it would have been in a simple ELS model of galaxy formation.

\section{CONSTRAINTS FROM EXISTING OBSERVATIONS}

 An array of observations are now available that provide
information, either direct or indirect, on galaxy formation
and structure formation in the early Universe.
Indirect observations include the large angular-scale polarization in the
cosmic microwave background (CMB), recently measured by {\it WMAP},
as well as the amplitude and fluctuations
of the cosmic near-infrared background (CIB). We will discuss these later,
and focus here on the search for discrete sources at the highest redshifts.

Available large telescopes, in space and on the ground, 
are capable of taking images of distant 
galaxies and/or obtaining spectroscopic data, reaching all the way to the
currently highest-redshift galaxy at $z=8.6$ (Lehnert et al. 2010).
There are two main techniques to locate $z > 6$ galaxies, both based
on the spectral imprint of hydrogen. In the first case, broad-band
photometry aims at identifying absorption breaks due to neutral
hydrogen in the vicinity of the source, and in the second case, narrow-band
techniques target the strong emission in the Lyman-$\alpha$ line (Stiavelli 2009).

\subsection{High-redshift Dropout galaxies}

\begin{figure}[ht]
\begin{center}
\includegraphics[width = 5 in] {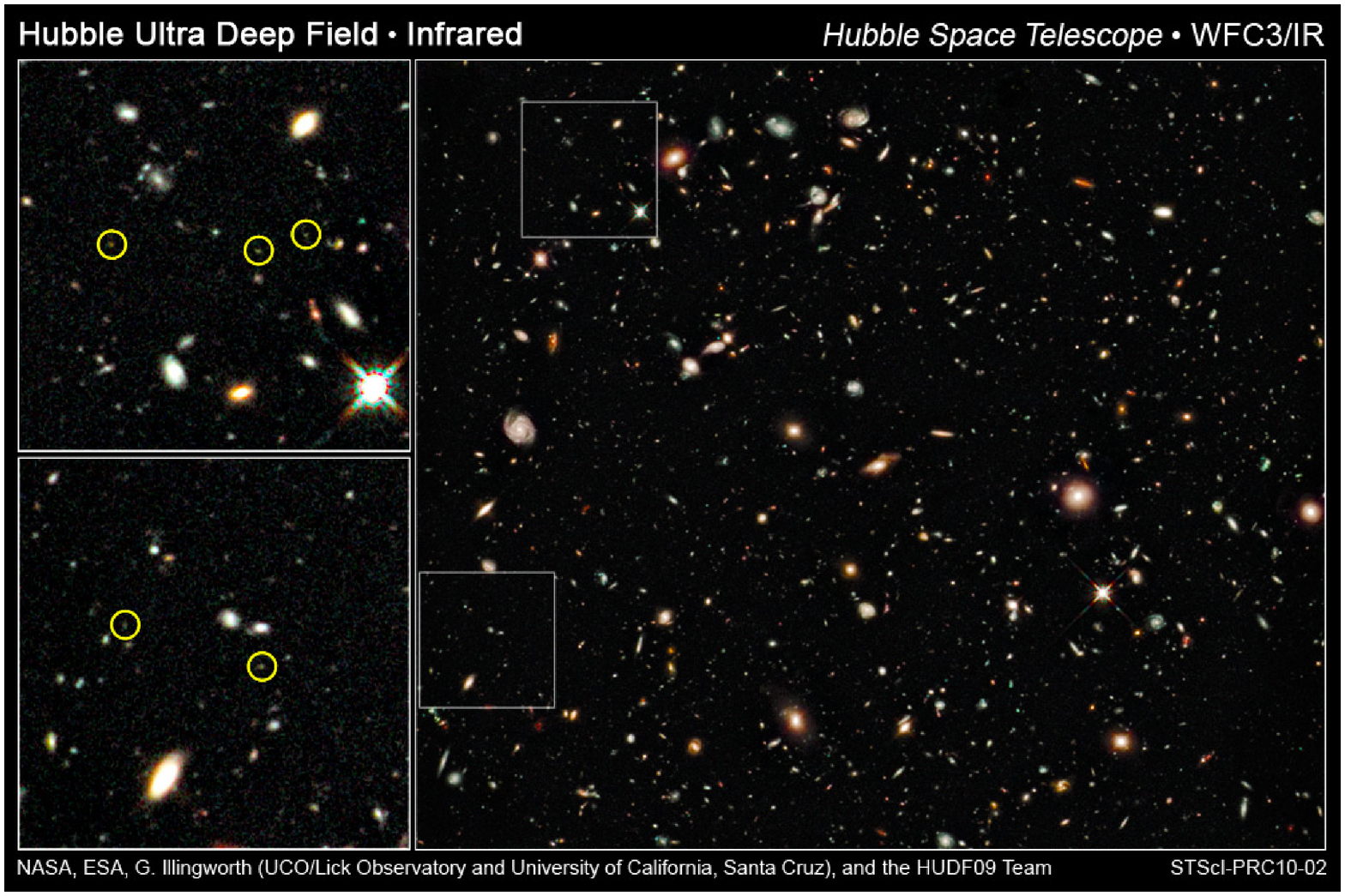}
\caption{Early galaxies in HST's deepest view of the Universe.
The image was taken with the newly installed WFC3/IR camera, with
the positions of newly discovered galaxies at $z\simeq 7-8$ indicated
by the circles in the zooms on the left-hand side. Figure courtesy of NASA.}
\label{fig_HUDF}
\end{center}
\end{figure}

Utilizing the exquisite near-IR sensitivity of the newly installed
Wide Field Camera~3 (WFC3) on board the HST,
deep images of the Hubble Ultra Deep Field (HUDF) and other fields opened up
an unprecedented window into the distant Universe (see {\bf Figure \ref{fig_HUDF}}).
High-redshift galaxies were identified by the so-called dropout technique,
using multi-band imaging (for a recent review, see Robertson et al. 2010).

\begin{figure}[ht]
\begin{center}
\includegraphics[width = 5 in] {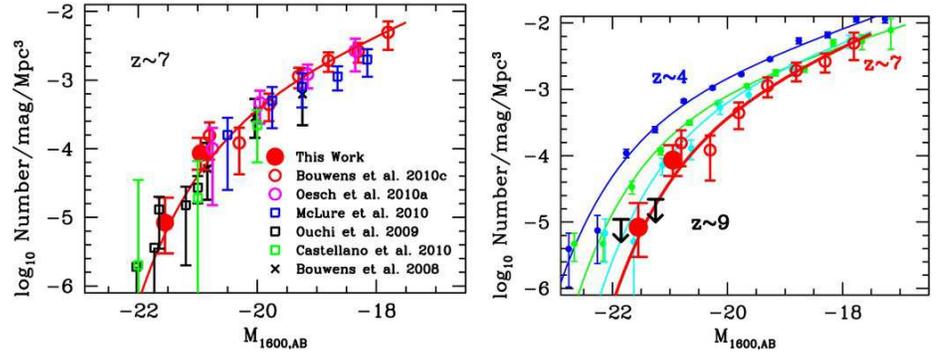}
\vspace{-2 in}
\caption{UV luminosity function at $z\sim 7$. Shown is the number density
of sources per unit magnitude vs. the absolute (soft) UV AB magnitude.
{\it Left panel:} LF at $z\sim 7$, as derived from HST NICMOS and
ground-based observations (large solid red circles), together with other
determinations, as labelled in the figure.
Overplotted is the best-fit Schechter function (solid red line).
{\it Right panel:} A comparison of the UV LF at $z\sim 7$ (solid red circles),
with those at $z\sim 6$ (cyan), $z\sim 5$ (green), and $z\sim 4$ (blue).
Evidently the LF evolves over the redshift interval considered here.
Adopted from Bouwens et al. (2010a).}
\label{UV_LF}
\end{center}
\end{figure}

The galaxy luminosity function (LF) at $z\sim 7$ was
derived from the combined observations by HST
and large ground-based telescopes
(Bouwens et al. 2010c; Ouchi et al. 2009; Castellano et al. 2010).
Wide-field observations using the ground-based telescopes
are important to determine the bright-end of the LF,
whereas HST is able to detect fainter galaxies. 
The LF is fit by a Schechter function 
which is described by a power-law towards the
faint end such that $\propto L^{-\alpha}$ (see {\bf Figure~\ref{UV_LF}}). 
The faint-end slope is a critical quantity to
derive the global star formation rate density
and to estimate the ionizing photon budget for hydrogen reionization
(Stiavelli, Fall \& Panagia 2004).
Athough the current data generally suggests a steep power-law 
with $\alpha \sim 1.7-1.9$,
it does not yet allow researchers to make a precise determination of $\alpha$.
More data to be aquired 
by the Cosmic Assembly Near-Infrared Deep Extragalactic Legacy
Survey\footnote{http://candels.ucolick.org/} (CANDELS)
using HST will reduce the uncertainty in the faint-end
slope substantially.

The newly-discovered galaxies beyond $z\sim 7$ appear to be
quite ``blue''. It is convenient to characterize a galaxy's 
stellar population by the UV spectral slope, $\beta$, where 
the flux density is: $f_{\lambda}\propto \lambda^{\beta}$.
The $z\sim 7$ galaxies show an unusually hard UV continuum 
with $\beta < -3$,  with fainter sources having bluer continuum
(Bouwens et al. 2010b,c; Finkelstein et al. 2010).
This is in pronounced contrast with local starburst galaxies
and Lyman-break galaxies at $z<6$ that have typically $\beta \sim -2$.
Interestingly, the steep continua of the $z\sim 7$ galaxies 
can be accounted for by stars with very low metallicities,
$Z < 5\times 10^{-4} Z_{\odot}$ (Taniguchi, Shioya \& Trump 2010).

\begin{figure}[ht]
\begin{center}
\includegraphics[width = 5 in] {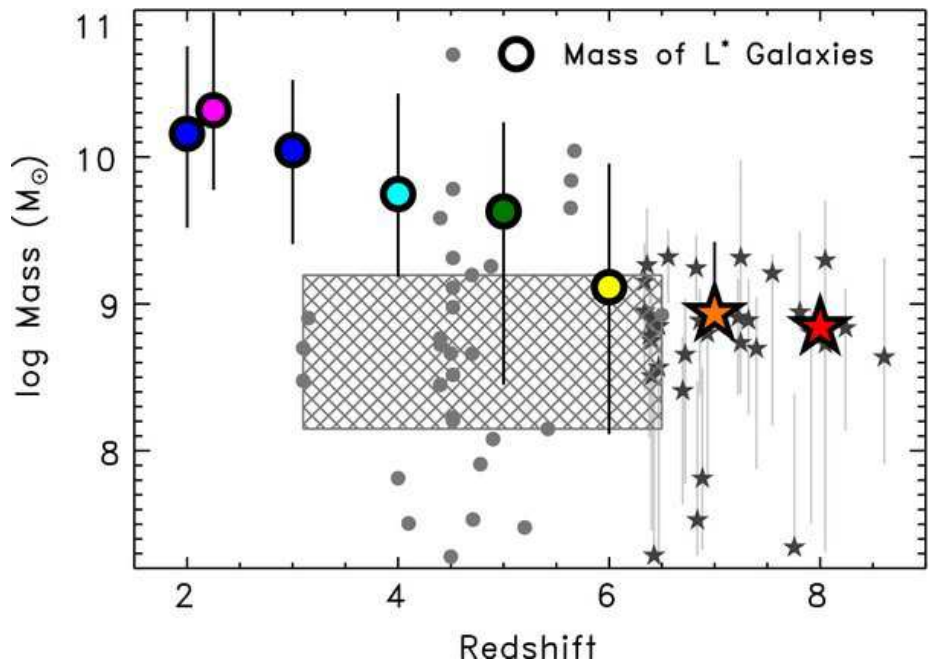}
\vspace{-1.5 in}
\caption{Stellar mass of high-redshift galaxies.
The colored symbols represent data for LBGs with characteristic luminosity
($L^{\ast}$). It is evident that stellar masses in typical LBGs decreases
with redshift. The small grey circles denote LAEs for comparison, and the
grey hatched region shows the interquartile range. The highest redshift
LBGs seem to be more similar to the LAEs than to LBGs at lower redshift.
Adopted from Finkelstein et al. (2010), where all references for the data shown here can be found.}
\label{Star_mass}
\end{center}
\end{figure}

The star formation history of individual galaxies
can be inferred from the mass density of long-lived stars.
Infrared observations by the {\it Spitzer Space Telescope} provided
information on the color, or shape of the spectral energy distribution (SED),
of high-$z$ galaxies. The data have been used to estimate the stellar mass
and approximate star formation histories of those galaxies
(Eyles et al. 2007; Stark et al. 2007; Labbe et al. 2010).
Luminous $z\sim 7$ galaxies have stellar masses of $10^{9-10}$
solar masses (see {\bf Figure~\ref{Star_mass}}). Obviously these luminous galaxies 
are not the first galaxies of our definition, but likely are {\it descendants} 
of the first galaxies.
A sample of $z\sim 7$ galaxies shows evidence
of extended star formation over a mean period
of 300 Myr (Gonzalez et al. 2010). This is indicative that star formation in
these galaxies and their progenitors must have begun at redshifts $z>10$ 
(Mobasher et al. 2005; Wiklind et al. 2008).
The $z\sim 7$ galaxies may thus have preserved the signature of star 
and galaxy formation in the pre-reionization era.

\subsection{Lyman-$\alpha$ Emitters}

There is another population of high-redshift galaxies,
characterized by strong Lyman-$\alpha$ line emission.
The LF of the Lyman-$\alpha$ emitters (LAEs) has been
obtained by Hu et al. (2004) and by Malhotra \& Rhoads (2004),
and more recently by observations with the
Subaru telescope (Ouchi et al. 2010; Hu et al. 2010).

\begin{figure}[ht]
\begin{center}
\includegraphics[width = 5 in] {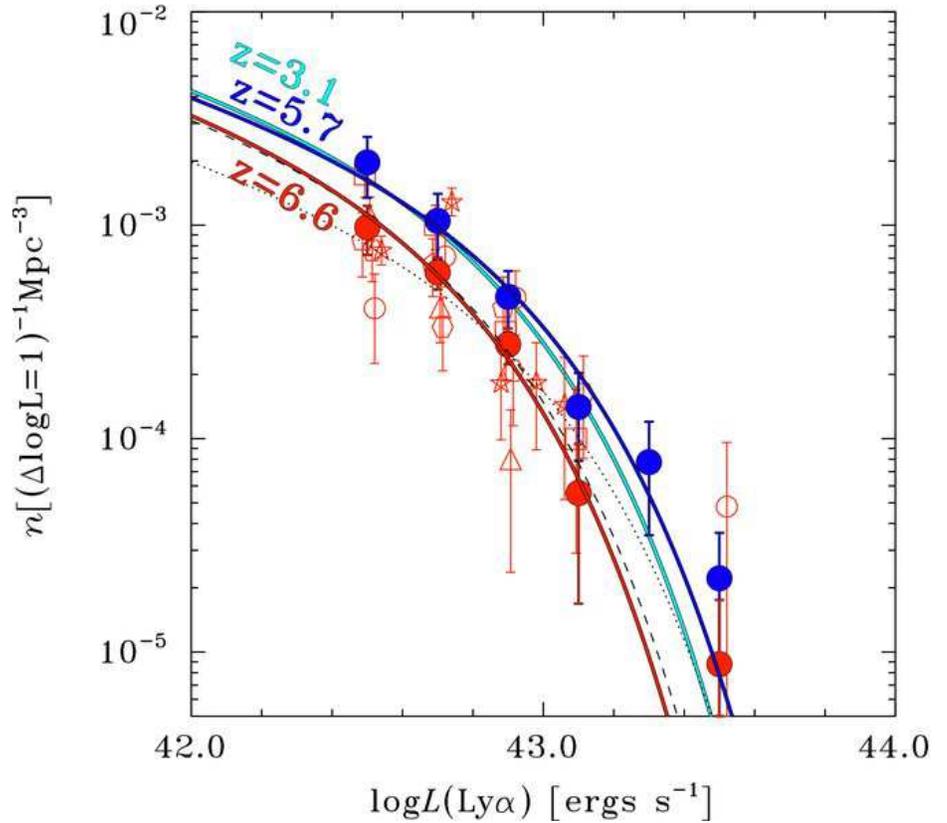}
\vspace{-1.3 in}
\caption{Evolution of Lyman-$\alpha$ luminosity function.
Shown is the number density of LAEs vs. Lyman-$\alpha$ luminosity for
three different redshifts, as labelled in the plot. The $z=6.6$ data
was derived from the 1~deg$^2$ wide Subaru/XMM-Newton Deep Survey (SXDS)
field, to be compared with previous measurements of the LF at lower redshifts.
Solid lines give various fits to the Schechter function. It is evident
that there is very little evolution from $z=3.1$ ({\it cyan solid line}) to
$z=5.7$ ({\it blue filled circles and solid line}), but significant
evolution towards $z=6.6$ ({\it red filled circles and solid line}).
The open symbols show the less precise results from smaller, 0.2~deg$^2$,
fields, which cannot reliably establish whether evolution is present or not.
This demonstrates the need for wide-field surveys to measure high-$z$
LFs with the required precision.
Adopted from Ouchi et al. (2010).}
\label{LF_EVOL}
\end{center}
\end{figure}

The evolution of the LAE LF across a redshift of six
is particularly interesting because the observed 
Lyman-$\alpha$ luminosity of a galaxy can be
significantly affected by neutral hydrogen in the IGM 
(see {\bf Figure~\ref{LF_EVOL}}).
The evolution of the IGM density and the neutral fraction can be
imprinted in the apparent LF
of LAEs (Dijkstra, Wyithe \& Haiman 2007; Iliev et al. 2008; Dayal,
Ferrara \& Gallerani 2008). Even the local and large-scale velocity field 
of the IGM affect the line profile and luminosity of 
individual galaxies (Dijkstra \& Wyithe 2010; Zheng et al. 2010).
Recent observations by Kashikawa et al. (2006),
Ouchi et al. (2009) and Hu et al. (2010)
showed that the LF evolution from z=5.7 to z=7
is small. The abundance of LAEs decreases at $z > 5.7$, 
indicating either that there was a slight
change in the neutral fration of the IGM over the time,
or that the galaxies themselves evolved.
Because of radiative transfer effects
for Lyman-$\alpha$ photons with their large
absorption and scattering cross-sections, 
the relationship between the intrinsic Lyman-$\alpha$ 
luminosity and the apparent, i.e., observed, one
is rather involved. 
The appearance of LAEs depends on density and velocity 
structures of the IGM surrounding them (McQuinn et al. 2007; 
Zheng et al. 2010).
LAEs themselves could be sources of 
reionization, which decrease the neutral fraction
of the IGM in their vicinities.
Interpreting the LF function evolution is thus difficult.
Large-volume cosmological simulations with Lyman-$\alpha$ radiative
transfer will be needed to quantify and more fully understand 
this complex interplay.

An important question is what the dominant sources
of reionization are.
The available observations robustly show that
the currently probed high-redshift galaxies, presumably the most luminous ones 
at the respective epochs, are not the dominant sources
of reionization. This is evident by simply counting
the total number of ionizing photons from the observed 
galaxies, and comparing them with the 
critical ionizing photon production rate for reionization
(Madau, Haardt \& Rees 1999; Robertson et al. 2010).  
There must have been many more faint galaxies that contributed
as reionization sources. Interestingly, even if one
integrates the currently estimated LF
to the very faint end, the estimated ionizing photon budget
still falls short of what is required to reionize the
Universe (Ouchi et al. 2009). 
Apparently, faint galaxies must have had large photon escape fractions,
and/or harbored stars with a more top-heavy IMF. 
Alternatively, there may have been
different types of early sources of reionization, such as mini-quasars
and massive Pop~III stars (Ricotti \& Ostriker 2004;
Sokasian et al. 2004; Kuhlen \& Madau 2005).

\section{THEORETICAL STUDIES} 

\subsection{Overview} 

The formation of the first galaxies is an intrinsically more complex
process, compared to the appealing simplicity of how the first stars
formed. In the latter case, the initial conditions are cosmologically
determined, and the relevant physical processes are virtually all known
(e.g., Yoshida, Omukai \& Hernquist 2008). 
In the standard hierarchical ($\Lambda$CDM) structure 
formation model, the first generation of stars is formed before galaxies emerged.
Feedback effects from these stars are thus expected to 
play a key role in setting the scene, {\it i.e., the initial conditions}, 
for first galaxy formation (see {\bf Figure~\ref{ASSEMBLY}}). In turn, the nature
of the first stars may be imprinted in various 
properties of the first galaxies.

\subsubsection{Formation Epoch} 
When did the first galaxies form? This is an intricate question,
because it is directly related to the definition of `first galaxy',
as discussed in Section~2.  
If minihalos were the hosts of the first galaxies 
(Ricotti, Gnedin, \& Shull 2002a, 2002b, 2008),
the very first galaxies would be formed at $z > 40$ 
within standard $\Lambda$CDM cosmology 
(Miralda-Escud\'{e} 2003; Naoz, Noter \& Barkana 2006).
However, it is more plausible that continuous star-formation 
can be sustained in larger mass dark matter halos, where
at least atomic hydrogen cooling operates efficiently.
Such large halos with virial temperatures greater than $\sim 10^{4}$\,K
are significantly biased objects at $z > 15$ (Miralda-Escud\'{e} 2003; 
Gao et al. 2007). The abundance of the rare density peaks
sensitively depends on the assumed cosmological parameters,
most notably on the fluctuation amplitude of the initial 
density field at the relevant mass (length) scales.
Typically, such atomic cooling halos, corresponding to $\sim 2\sigma$-peaks in
the Gaussian random field of initial density perturbations, are predicted
to form at $z\sim 10-15$, or roughly 500~Myr after the Big Bang. Thus, the
epoch of the first galaxies lies just beyond the current horizon of
observability, and the {\it JWST} or the next-generation, 30-40m, ground-based
telescopes will be able to detect them.

\subsubsection{Stellar Feedback} 
A key element in the physics of first galaxy formation is the feedback
from the first stars, and the complications arising from it.
If the first stars were massive, they would have exerted a strong influence
on the gas in the host halo by injecting significant
energy either by radiation or by supernova explosions. 
Then the next episode of star formation was likely to be
delayed for a long time, comparable to the dynamical time
of the massive halo, even if the halo's virial temperature
well exceeds $10^{4-5}$ K. Specifically,
delay times of a few $10^7$\,yr are predicted, which corresponds to
a significant fraction of the Hubble time at $z\sim 15$.
Cosmological simulations performed so far generally support
the notion (Johnson \& Bromm 2007; Yoshida et al. 2007a;
Alvarez, Wise \& Abel 2009). The strength of the feedback effect
could in principle be reflected in the very faint-end
shape of the luminosity function of high-redshift $(z>7)$ galaxies
(Haiman 2009). The characteristic mass of the first stars,
ultimately driving the strength of the negative feedback, may thus be constrained.

\subsubsection{Conditions for Star Formation} 
\begin{figure}[ht]
\begin{center}
\includegraphics[width = 5 in] {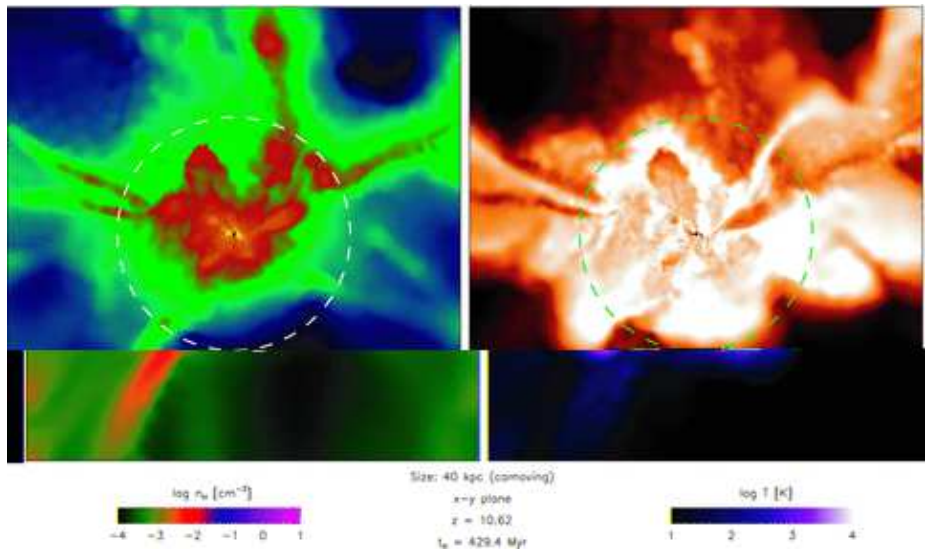}
\vspace{-2 in}
\caption{Turbulent collapse into the first galaxy.
Shown is the hydrogen number density
({\it left-hand panel}) and temperature ({\it right-hand panel}) in the inner 4\,kpc (physical), surrounding the BH at the center of the galaxy, indicated by the filled black circle. The dashed lines
denote the virial radius at a distance of 1 kpc. Hot accretion dominates where gas is accreted directly from the IGM and shock-heated to $10^4$\,K. In contrast,
cold accretion becomes important as soon as gas cools in filaments and flows towards the center of the galaxy. These cold streams
drive a prodigious amount of turbulence and create transitory density perturbations that could in principle become Jeans-unstable.
Adopted from Greif et al. (2010).}
\label{TURB}
\end{center}
\end{figure}

When we approach the assembly of the first galaxies, the degree of complexity
is greatly enhanced compared to the simplicity that governed the formation
of the very first stars. In particular, this emerging complexity set the stage
for the second generation of star formation that occurred inside the first
galaxies. The existence of heavy elements, and possibly of dust grains, the
degree of turbulence, and the likely presence of dynamically significant
magnetic fields all need to be taken into account.
External radiation fields, either from nearby stars and galaxies, or built up
globally, also regulated the formation of molecular gas clouds (Ahn \& Shapiro 2007;
Johnson, Greif \& Bromm 2007; Susa 2008).
Star formation in the first galaxies is thus 
as complicated as present-day star formation, 
and may also be qualitatively similar.
 Recent cosmological hydrodynamical simulations confirmed that
strong turbulence develops within large, proto-galactic halos
(Wise \& Abel 2007; Greif et al. 2008).
Turbulence is generated by supernova explosions or dynamically
through dark matter halo mergers, or more generally as a result of gravity-driven
virialization. The turbulence is typically supersonic, related to the cold-flow
accretion streams that feed gas into the very centers
of the first galaxies (see {\bf Figure~\ref{TURB}}).
In the presence of rapid cooling by atomic hydrogen and by heavier atoms such as
carbon, oxygen and iron, the turbulent gas might settle into rotationally supported,
central disks (Wise \& Abel 2007). We thus have obtained a much improved
picture of the physical conditions just prior to the onset of the
initial starburst inside the first galaxies.

\subsubsection{Simulated versus observed galaxies} 
Current state-of-the-art cosmological simulations followed the formation 
of objects with still rather low masses,
typically $\sim 10^{8} M_{\odot}$.
The real target of the next-generation telescopes, however, will 
be those with masses $\gtrsim 10^{9} M_{\odot}$ (Mashchenko, Wadsley \&
Couchman 2008; Pawlik, Milosavljevic \& Bromm 2011). Therefore, there still
remains a large gap between the available highly-resolved, ab initio simulations
and the realistic targets for the upcoming observations. For the simulation
community, much work is still required in building the bridge to the
observations. We already know the rough outlines of the $\gtrsim 10^9 M_{\odot}$
halo formation problem though.
Semi-analyic models of galaxy formation
combined with large-volume cosmological simulations
show that such ``luminous'' galaxies appear as early as
$z\sim 15-20$ (Springel et al. 2005; Lacey et al. 2010). 
A concerted use of both of these approaches,
semi-analytical and ab-initio simulations, will be needed to address the many
important questions about the formation of the first galaxies
(Benson 2010; Raicevic, Theuns \& Lacey 2011).

\subsection{Pre-Galactic Metal Enrichment}
The first galaxies are plausible sources of heavy elements
that existed in the IGM at high redshifts
(Songaila et al. 2001; Simcoe 2006; Ryan-Weber et al. 2009).
The IGM metalicity evolution can place constraints on the prior 
star formation history (Maio et al. 2011). 
Although it has also been proposed that
Pop~III stars, formed in minihalos, can contribute
to early chemical evolution (Yoshida et al. 2004; 
Tornatore, Ferrara \& Schneider 2007;
Greif et al. 2007),
recent observations suggest that the C\,IV abundance declines at $z>6$  
(Becker, Rauch \& Sargent 2009).
It is therefore likely that the heavy elements were dispersed
by galaxies that had formed around $z\sim 6$.
Assuming that the first galaxies are the 
dominant source of the IGM enrichment {\it and} reionization,
one should be able to build a consistent model for 
the reionization history, the galaxy luminosity function
and its evolution, as well as the stellar population and chemical evolution
in the first galaxies (Choudhury \& Ferrara 2006).

Galactic scale outflows driven by radiation pressure
from hot stars and/or by supernovae can transport
heavy elements into the IGM
(Madau, Ferrara \& Rees 2001; Mori, Ferrara \& Madau 2002; Wada \& Venkatesan 2003).
How exactly this happened during the reionization epoch can be inferred by comparing
the metallicity evolution and the star-formation history.
The currently available data seem to point to 
delayed enrichment via galactic outflows, rather 
than prompt enrichment (Kramer, Haiman \& Madau 2010).
Three-dimensional cosmological simulations consistently
show that the IGM metal pollution is patchy, leaving
a large volume of unpolluted, chemically pristine 
regions at $z > 6$ (Bertone, Stoehr \& White 2005; 
Tornatore, Ferrara \& Schneider 2007).
One possible implication of such inhomogeneous enrichment is the existence
of Pop~III star clusters or SN explosions
at lower redshifts, $z < 6$ (Scannapieco et al. 2005; Johnson 2010).
Such objects, if they existed, would be an exciting
target for direct observations with the {\it JWST} and
future 30-40m ground-based telescopes.

\begin{figure}[ht]
\begin{center}
\includegraphics[width = 5 in] {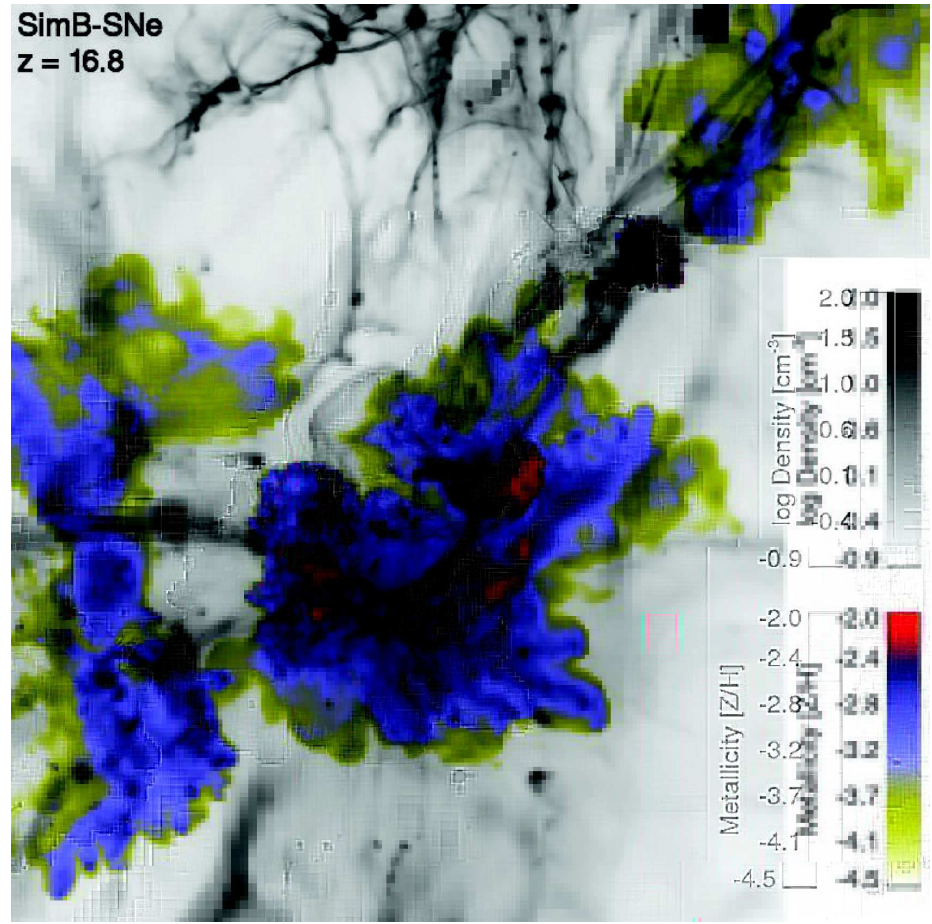}
\vspace{-1. in}
\caption{Metal enrichment in the first galaxy.
Shown is the aftermath of tens of pair-instability supernovae (PISNe) which exploded inside
the progenitor minihalos. The situation here corresponds to $z\simeq 17$. The projection of metallicity is shown in
color, and that of gas density in shades of grey, with values indicated
by the insets. The box has a proper size of 8.6~kpc.
Adopted from Wise \& Abel (2008).}
\label{SN}
\end{center}
\end{figure}

Inside individual first galaxies, the mixing of heavy elements
can occur rapidly. Hydrodynamic simulations confirmed this,
showing that a large volume of the halo gas in the first galaxies
is already metal-enriched before it condenses again 
to trigger the next episode of star formation (Greif et al. 2010; 
JH Wise et al., submitted). Specifically, metallicities inside the first
galaxies prior the the initial starburst can reach average levels
of already $\sim 10^{-3} Z_{\odot}$, with maximum levels even up to
an order of magnitude higher (see {\bf Figure~\ref{SN}}).
The degree of mixing and details of the chemical enrichment
history can be studied by the very promising approach of stellar
archaeology (Section~7). In particular, the metallicity distribution and
the relative elemental abundance patterns of stars in 
dwarf galaxies in the Local Group may preserve
the fossil record of early chemical enrichment.

\subsection{Star Formation in the First Galaxies}
 Outstanding questions regarding star-formation in the first galaxies
are the star-formation efficiency, the stellar IMF, 
and the strength of stellar feedback. These three elements
are indeed closely connected to each other. 
The star-formation efficiency is largely affected by the ability
of the halo gas to cool and condense. Because the gas density is low
initially, cooling by atomic heavy elements such as carbon,
oxygen, and iron is effective in early phases 
(Bromm et al. 2001; Bromm \& Loeb 2003a; Santoro \& Shull 2006; 
Omukai et al. 2005; Maio et al. 2010).
Unlike hydrogen molecules, which are fragile to soft UV radiation
in the Lyman-Werner (LW) bands, cooling by metallic
atoms and ions can operate even under the 
influence of a UV radiation field (Maio et al. 2007; Safranek-Shrader, Bromm \& Milosavljevic 2010).

 The stellar IMF is more difficult to address. Observationally,
at least for local star-forming regions, it is well determined 
to peak at roughly solar masses and to exhibit a power-law 
extension towards higher masses: $dN/d\ln M \propto M^{x}$ with
$x \sim -1.35$ (Salpeter 1955; Zinnecker \& Yorke 2007).
However the mechanism that shapes the IMF is not well 
understood even in the local Universe. 
It is often thought that predicting the IMF for the first stars 
would be simpler in many ways, and that it would be 
more top-heavy, with stars more massive than a few 
tens of solar masses being predominant (for a review of the argument, see Bromm \& Larson 2004). 
In the first galaxies, there are a number of physical 
ingredients that have been suggested to significantly
affect the IMF: supersonic turbulence (Wise \& Abel 2008; Greif et al. 2008; 2010), 
atomic cooling by heavy elements (Bromm et al. 2001; Santoro \& Shull 2006; 
Smith, Sigurdsson \& Abel 2008),
cooling by dust (Schneider et al. 2006; Omukai et al. 2005),
the angular momentum transfer and heating due to 
magnetic fields (Schleicher et al. 2010), the initial degree of ionization
(Nagakura \& Omukai 2005; Johnson \& Bromm 2006; Yoshida, Omukai \& Hernquist 2007;
Cazaux \& Spaans 2009), and a lower floor to the attainable gas temperature
set by the CMB (Larson 1998; Schneider \& Omukai 2010). 
All these processes acted to render
star formation in the first galaxies similar again to the present-day case.
In particular, the presence of supersonic turbulence likely 
allowed the formation of multiple stars in a molecular cloud,  
with a broad mass spectrum that may have resembled the
local self-similar form towards high masses. 
This expectation, however,
still needs to be tested with sophisticated simulations.

The ionization degree is important particularly for a primordial gas.
The IGM can be ionized by radiation from the first stars 
(Kitayama et al. 2004; Whalen et al. 2004), 
blastwaves driven by the first supernovae (Bromm et al. 2003;
Machida et al. 2005), by cosmic rays (Vasiliev \& Shchekinov 2006;
Jasche, Ciardi \& Ensslin 2007; Stacy \& Bromm 2007), 
by X-rays emitted from early mini-quasars (Oh 2001;
Ricotti \& Ostriker 2004; Kuhlen \& Madau 2005),
or through dark matter annihilation/decay 
(Ripamonti, Mappeli \& Ferrara 2007; Iocco et al. 2008; Spolyar, Freese \& Gondolo 2008).
An initially ionized gas of primordial composition can cool 
to $\sim 100$~K,
where cooling by hydrogen deuteride (HD) molecules becomes important.
The abundance of additional free electrons here catalyzes a boost
in H$_{2}$ formation, which in turn leads to the build-up of a critical
abundance of HD, thus enabling this low-temperature cooling channel 
(e.g., Johnson \& Bromm 2006).
Primordial stars formed under this condition, the so-called
Population III.2 stars (McKee \& Tan 2008; Bromm et al. 2009),
are thought to include ordinary massive stars 
(Johnson \& Bromm 2006; Yoshida, Omukai \& Hernquist 2007; Clark et al. 2011).
However the relative importance of Pop~III.2 stars
remains uncertain (Trenti \& Stiavelli 2009;
Wolcott-Green \& Haiman 2010).
If Pop~III.1 and Pop~III.2 stars have different characteristic masses,
detection of high-redshift supernovae of
different types, pair-instability SNe and core collapse SNe,
will provide constraints on the relative formation rates of
PopIII.1 and PopIII.2 stars.

\begin{figure}[ht]
\begin{center}
\includegraphics[width = 5 in] {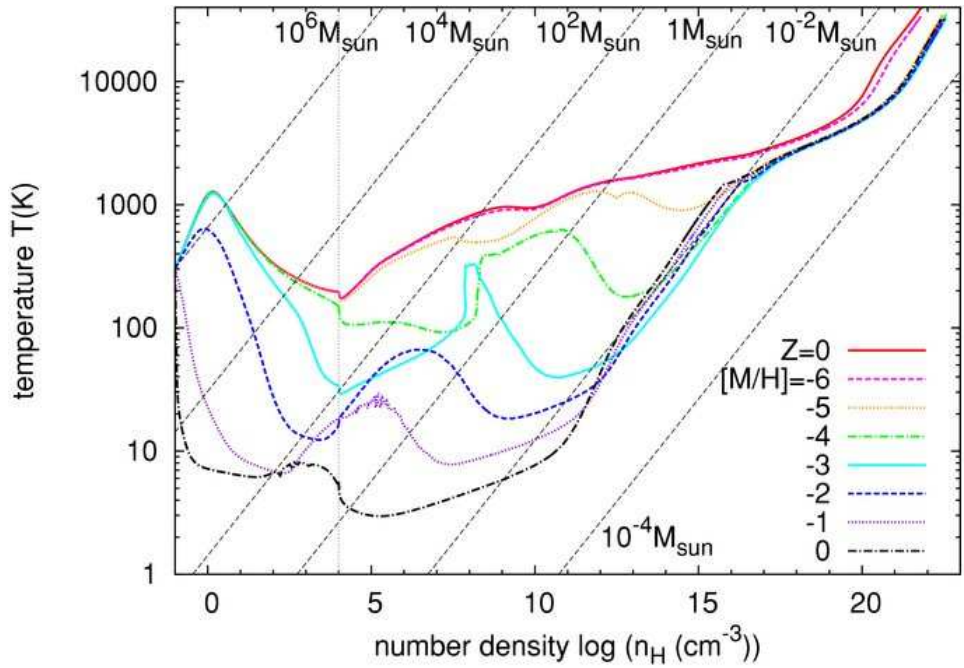}
\caption{Thermal evolution of pre-stellar gas with various metallicities.
The constant Jeans masses are indicated by the dashed lines.
Characteristic temperature dips are caused by cooling due to
atomic line cooling at low densities, molecular cooling at intermediate 
densities, and dust thermal emission at high densities.
Adopted from Omukai, Hosokawa \& Yoshida (2010).}
\label{fig_lowZ}
\end{center}
\end{figure}

Because of the chemical feedback discussed in Section 4.2,
many stars in the first galaxies
are probably metal enriched.
Detailed calculations on the thermal evolution of a low-metallicity
gas have been carried out (Schneider et al. 2002; 
Jappsen et al. 2007; Omukai, Hosokawa \& Yoshida 2010). 
The results suggest that dust thermal emission remains
an efficient cooling mechanism up to very high densities 
where atomic line cooling is ineffective.
Dust cooling allows fragment masses to reach very small,
essentially opacity-limited values of $\gtrsim 10^{-2}M_{\odot}$
(see {\bf Figure~\ref{fig_lowZ}}).
Three-dimensional simulations are needed to determine the
ability of a low-metallicity gas to fragment, and to follow the subsequent
accretion and merging history of the growing protostars.
One such study has been carried out by Clark, Glover \& Klessen (2008)
who employed a tabulated barotropic equation of state
for a low-metallicity gas. The challenge now is to extend such calculations  
to realistic initial conditions, and to self-consistently determine the
equation of state during the dynamical collapse.

\subsection{Radiation from the First Galaxies}

\subsubsection{Ionizing photon budget and the escape fraction}

First galaxies are promising sources of ultra-violet photons
that reionized the intergalactic hydrogen.
A critical quantity is the escape fraction of ionizing photons, 
$f_{\rm esc}$.
Recent simulations that couple
the hydrodynamics of the gas in the vicinity of the central star
cluster to the continuum radiative transfer of the ionizing radiation
from these stars find that the escape fraction strongly evolves with
time (Johnson et al. 2009). Initial values are close to zero, when
gas densities are still high, and most of the ionizing radiation is bottled up
inside the galaxy. With time, however, the photo-ionization heating creates
a central high-pressure bubble which in turn drives a strong outflow. Densities
thus decrease, until ionizing photons can freely escape into the IGM, leading
to a large instantaneous escape fraction of 
$f_{\rm esc}\sim 1$. Time-averaged escape fractions are typically quite
large, $f_{\rm esc}\sim 0.1 - 0.8$ (Wise \& Cen 2009; 
Razoumov \& Sommer-Larsen 2010). 
Extinction by a substantial amount of dust can reduce it
to $f_{\rm esc} \sim 0.1$ or less (Gnedin, Kravtsov \& Chen 2008; Yajima et al. 2009).
Available observations suggest $f_{\rm esc} < 0.01$ for 
low-redshift galaxies (e.g., Bridge et al. 2010), 
whereas $f_{\rm esc} = 0.01 - 0.1$
for $z\sim 1-3$ galaxies (Inoue et al. 2006;
Shapley et al. 2006; Siana et al. 2007; Iwata et al. 2009). 
There are indirect hints from observations of high-redshift galaxies
regarding the escape of ionizing radiations, and the stellar populations
responsible for this emission (e.g., Jimenez \& Haiman 2006).

The ionizing photon budget derived from
the currently estimated UV luminosity function of $z>6$ galaxies
falls short of what is necessary to reionize the
Universe (Ouchi et al. 2009). A possible resolution may be
either that faint, low-mass galaxies host a substantially
``bluer'' stellar population, or that the escape fraction
from the faint galaxies is actually large. This interpretation
of the data agrees with the results from recent cosmological simulations,
which consistently predict such large values of $f_{\rm esc}$.

\subsubsection{Global signature}

The radiation produced by the first galaxies cumulatively contributes to
reionization, to the CIB, and to the redshifted 21-cm signal. Here, we only
briefly discuss these global signals, as they have been extensively
reviewed elsewhere: 21cm cosmology by Furlanetto, Oh \& Briggs (2006),
Barkana \& Loeb (2007),
and Morales \& Wyithe (2010), the CIB by Hauser \& Dwek (2001), Kashlinsky (2005),
and Arendt et al. (2010), and reionization in the review papers mentioned in
Section~1.

Cosmic reionization imprints distinct large angular-scale patterns in the CMB 
polarization maps. 
CMB photons are Thomson-scattered by free electrons in the reionized IGM. 
As a consequence, the CMB photons are polarized and the temperature fluctuations 
are damped. 
These signatures can be used to infer the approximate epoch of reionization. 
The seven-year {\it WMAP} data  
yields the CMB optical depth to Thomson scattering,
$\tau\simeq 0.09\pm 0.03$ (Komatsu et al. 2009), where
\begin{equation}
  \tau = \int_{0}^{z_{\rm reion}} {\rm d}\tau_{e} 
  \approx 0.0023\left [
    \left ( [1+z_{\rm reion}]^3 \, +\, 2.7\right )^{1/2}-1.93\right ].
\end{equation}
for the standard $\Lambda$CDM cosmology. 
Here, we have assumed for simplicity that the IGM is fully ionized at 
$z < z_{\rm reion}$.
The WMAP measurement provides an integral constraint on the total 
ionizing photon production at $z>6$.
The contribution from $z<6$ to the total optical depth
amounts only to $\tau \lesssim 0.04$ and thus a significant volume
fraction of the IGM must be ionized to $z=10$ or higher. Matching the
{\it WMAP} Thomson optical depth constraint provides a non-trivial
test for models of early star and galaxy formation. 
It is unlikely that reionization is completed very early
by massive Pop III stars (Cen 2003; Greif \& Bromm 2006; Haiman \& Bryan 2006).
More accurate polarization measurements by the {\it Planck Surveyor Satellite}
will further tighten the constraint on the Thomson optical depth, and in
addition might even allow researchers to estimate the reionization history of the
Universe (Holder et al. 2003; Mukherjee \& Liddle 2008).
The latter is usually expressed as the redshift-dependent free electron
fraction, $x_{\rm e} (z)$, which could be much more complex than the simple
step function, which is often assumed in approximate interpretations of the data
(see Fan, Carilli \& Keating 2006).

The first galaxies inevitably contributed to the CIB, through
the redshifted Lyman-$\alpha$ recombination line from the H{\sc ii} regions 
surrounding their stellar sources (Santos, Bromm \& Kamionkowski 2002;
Salvatera \& Ferrara 2003).
A vigorous debate has developed around the question of how important
still unresolved galaxies at the highest rdshifts are, compared to
more local, known sources (e.g., Kashlinsky et al. 2005; Thompson et al. 2007). 
If the difficult subtraction of foreground sources, such as the emission
from the interplanetary dust, can be reliably accomplished, a number of key 
parameters of the first galaxies might be derived from the CIB.
One is the typical mass of the first galaxies. In hierarchical structure
formation, the mass function is dominated by the lowest mass satisfying 
the first galaxy criteria (see Section~2). The corresponding
dark matter halos then exhibit clustering properties that are characteristic
for that mass scale. Those clustering properties are subsequently reflected
in the CIB fluctuation power spectrum
(Fernandez et al. 2010). A second quantity is the
escape fraction of hydrogen ionizing photons
from the first galaxies, which could possibly be inferred from 
the mean intensity of the CIB. The basic idea here is that the production
of rest-frame Lyman-$\alpha$ photons is greatly enhanced if the ionizing
radiation inside the first galaxies {\it cannot} escape into the IGM, where
densities are very low (recombination lines are emitted at a rate $\propto n^2$).
The measured CIB angular power spectrum can largely be
attributed to galaxies at $z<4$, but the possibility for
a contribution from $z>8$ sources still remains (Cooray et al. 2007).

Redshifted 21-cm emission from neutral hydrogen 
directly probes the topology of reionization 
(Furlanetto, Oh \& Briggs 2006). LOFAR
has already begun to collect data and is carrying out its initial calibrations.
It will provide statistical information on the distribution
of neutral hydrogen at $z\sim 6$, and will eventually 
be able to map out the distribution directly. Even more powerful is the planned
{\it Square Kilometer Array} (SKA), with an unprecedented sensitivity
and spectral coverage.
The clustering of the first galaxies can be used 
to study the topology of reionized regions.
If the first galaxies were dominant sources of 
reionization, their distribution should be anti-correlated
with ionized regions that appear as dark holes
in 21-cm maps (Lidz et al. 2009). 

These global signatures have the advantage that they do not suffer from 
incompleteness or selection effects of the target galaxies. Very small, 
faint galaxies that cannot be seen by JWST may in principle leave 
distinct signatures in the global quantities discussed here.

\section{THE FIRST SUPERMASSIVE BLACK HOLES}

The origin of SMBHs that power the luminous
quasars at high redshifts remains unknown.
Spectroscopic observations revealed that BHs with mass greater than 
10 billion solar masses were already in place when the age
of the universe was less than one billion years (for a review, see
Fan, Carilli \& Keating 2006). 
Potentially, the existence of such early SMBHs might pose 
a challenge to the current cosmological standard model which is based
on bottom-up, hierarchical structure formation.
The observed SMBHs have likely grown from some smaller seed
BHs that were formed earlier, in the progenitors of the luminous
quasar host. The first galaxies were plausible sites for seed BH formation,
but their own structure and evolution was likely affected by the presence
of such early BHs as well. We thus have to tackle a complex, feedback-regulated
problem, where our current knowledge is patchy at best. 

\begin{figure}[ht]
\begin{center}
\includegraphics[width = 5 in] {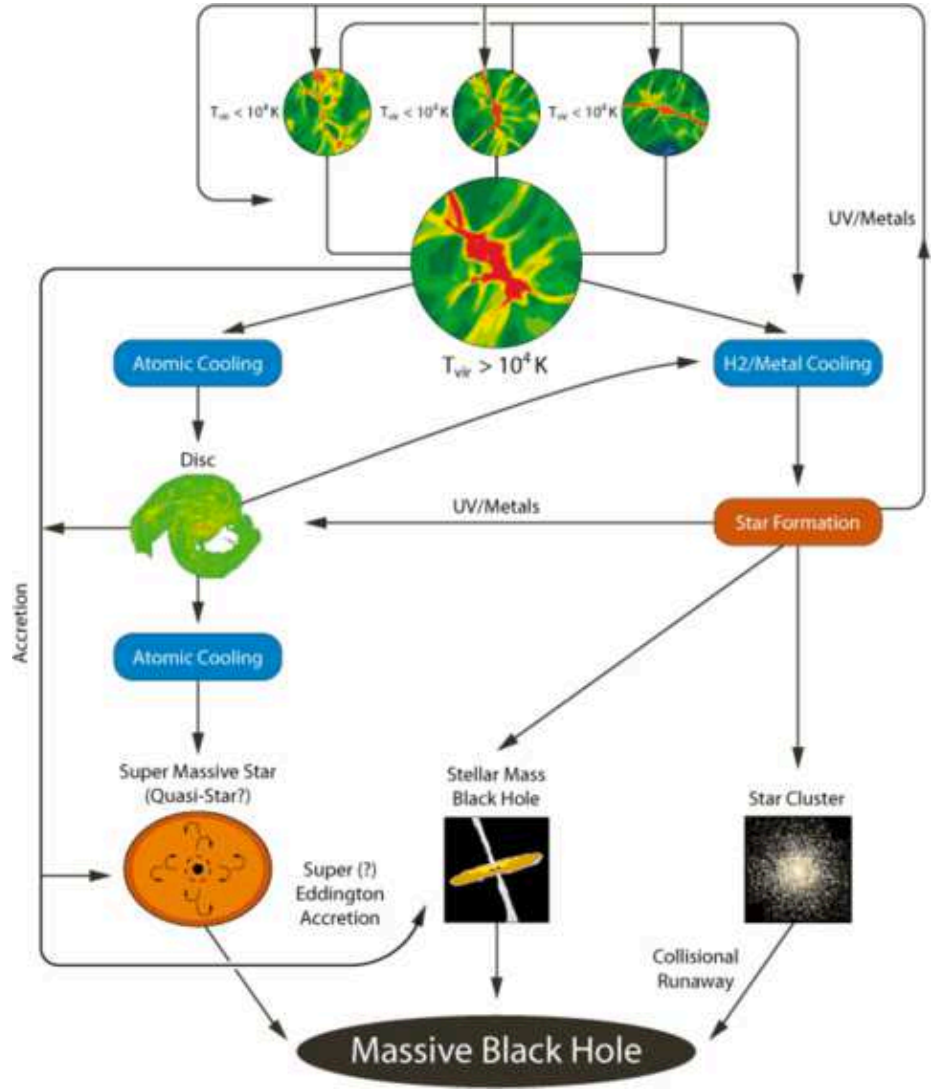}
\vspace{-0.5 in}
\caption{Pathways towards the first supermassive black holes.
Here, possible SMBH formation channels in high-redshift atomic cooling halos
are shown. The main bifurcation arises from whether the gas inside the first
galaxy can cool below $\sim 10^4$\,K, via H$_{2}$ or metal cooling, or not.
If the gas can cool, star formation will ensue. SMBH formation would then 
have to rely on stellar-dynamical processes of catastrophic runaway collisions. 
In the opposite case, the path towards a SMBH involves gas-dynamical processes, 
possibly resulting in the intermediate stage of a supermassive star (or quasi-star). 
Such a star would rapidly turn 
into a SMBH. 
Adopted from Regan \& Haehnelt (2009b).}
\label{PATH}
\end{center}
\end{figure}

It is instructive to consider a schematic representation of possible SMBH
formation pathways inside the first galaxies (see {\bf Figure~\ref{PATH}}).
Figure~\ref{PATH} is reproduced from Regan \& Haehnelt (2009b), who in turn adopt
the well-known flow-chart towards SMBH formation introduced by Rees (1984).
The key bifurcation concerns whether the gas inside the first galaxy, here
taken to be an atomic cooling halo, can cool below $\sim 10^4$\,K or not.
Such cooling depends on the presence of either H$_{2}$ or heavy-element
coolants. To prevent molecular hydrogen from forming, 
the presence of an extremely strong LW radiation background,
capable of photo-dissociating H$_{2}$ even in the presence of self shielding
would need to be invoked
(Bromm \& Loeb 2003b; Wise, Turk \& Abel 2008; Dijkstra et al. 2008;
O'Shea \& Norman 2008; Regan \& Haehnelt 2009a; Shang, Bryan \& Haiman 2010). 
To maintain
metal-free conditions in the first galaxies, star formation and SN activity
in the progenitor minihalos would have to be suppressed, which may be possible
in a subset of cases, in $\sim 10-20$\% of atomic cooling halos collapsing
at $z\gtrsim 10$ (Johnson et al. 2008). 
Below we discuss some of the SMBH
formation pathways in greater detail.

\subsection{Formation Models}

Devising viable models for SMBH formation has been a long-standing 
challenge in astrophysics (Rees 1984). The requirements on such models 
are even more stringent in the high-redshift case, where any formation 
channel has to operate on rapid timescales. There are currently two 
main ideas, one based on (Pop III) stellar seeds, some of them may 
grow via gas accretion and BH mergers, and one on the direct collapse 
of massive primordial gas clouds. Both classes of models face challenges, 
leaving still open the possibility for alternative, more exotic 
pathways towards SMBH formation. 

\subsubsection{Population III Stellar Remnants}
\begin{figure}[ht]
\begin{center}
\includegraphics[width = 5 in] {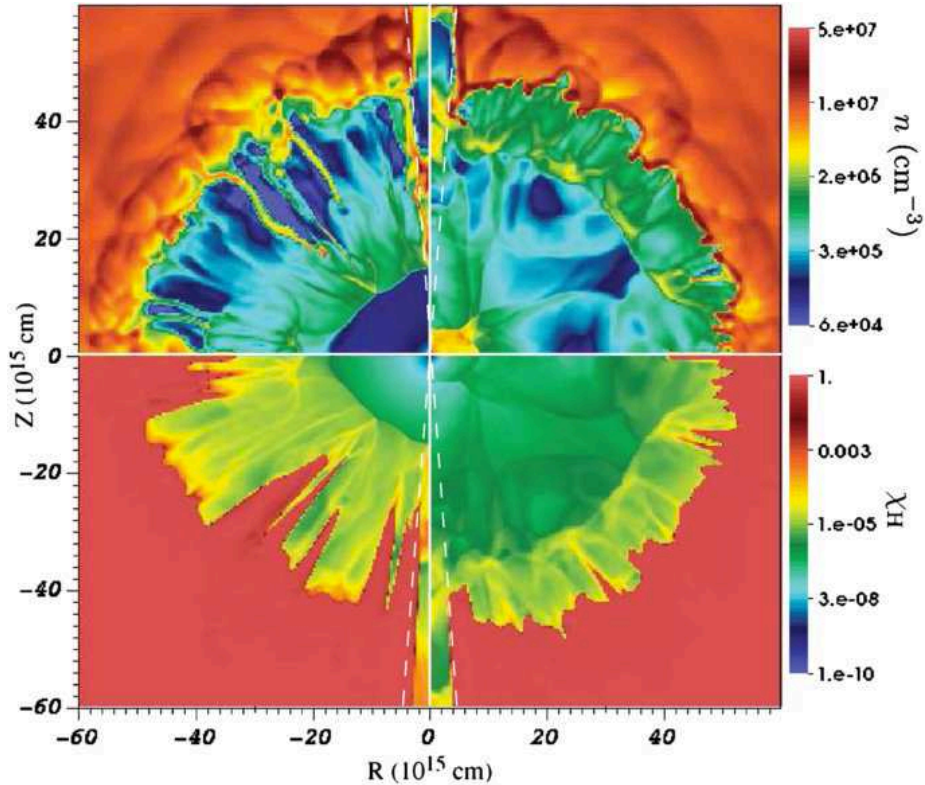}
\vspace{-1.0 in}
\caption{Accretion onto the first black holes.
Gas number densities $n$ ({\it top row}), and neutral fraction $\chi_{\rm H}$
({\it bottom row}) in the vicinity of the accreting black hole. Shown is
the situation during an accretion minimum ({\it left column}), and during
a maximum ({\it right column}). At maximum, central densities are high,
and the H\,II region grows in response. The structure near the vertical axis
({\it dashed lines}) is a numerical artifact. The resulting hydrodynamics
is complex, exhibiting overlapping in- and outflows that establish an
episodic pattern of accretion and radiation-pressure feedback.
Adopted from Milosavljevic, Couch \& Bromm (2009b).}
\label{BH_ACC}
\end{center}
\end{figure}

A popular model assumes that the remnant BHs of Pop~III
stars seeded the growth of SMBHs (Madau \& Rees 2001;
Li et al. 2007; Volonteri \& Rees 2006; Tanaka \& Haiman 2009).
In this case, the initial seed mass would be of order $100 M_{\odot}$.
Given efficient, Eddington-limited accretion, even such low-mass seeds
could readily grow to the SMBHs inferred to power the high-$z$ SDSS quasars
in the roughly 500\,Myr between seed formation and $z\sim 6$ (Haiman \& Loeb 2001).
Recent studies suggest, however, that the gas accretion onto 
early BHs is inefficient
until the BHs are incorporated into larger
mass halos. One impeding effect is
that the gas is already evacuated by photoionization heating
from the progenitor massive star (Kitayama et al. 2004; Whalen et al. 2004;
Alvarez, Bromm \& Shapiro 2006; Abel, Wise \& Bryan 2007). After the progenitor
star has died and directly collapsed into in intermediate mass BH, it thus finds
itself in a very low-density region. Accretion rates are then negligible for
at least the free-fall time of the dark matter host systems (Johnson \& Bromm 2007;
Pelupessy, Di Matteo \& Ciardi 2007; Alvarez, Wise \& Abel 2009). 
In addition, the radiative feedback from the accreting BH 
can reduce the cooling of the surrounding gas, e.g., by photo-dissociating
H$_{2}$, thus further reducing accretion. Even if
the gas supply in the vicinity of the remnant BH has been replenished, accretion
likely continues to be severely suppressed compared to the Eddington rate. This is
because of radiation pressure on the high-density infalling gas 
(Milosavljevic et al. 2009a,b). As a result, an episodic, quasi-periodic
accretion flow is established, with a time-average significantly below 
the Bondi-Hoyle and Eddington rates
(see {\bf Figure~\ref{BH_ACC}}).

This early bottleneck for growing the seeds to SMBHs
poses a serious challenge to the Pop~III stellar remnant scenario.
However, it is important to note that 
the emergence of SMBHs should not be too common, 
to be compatible with the abundance of observed luminous quasars
(Tanaka \& Haiman 2009). 
It is not necessary that a particular process is able 
to feed all seed BHs efficiently, although
there must be at least one physical mechanism that enables the 
early formation of SMBHs perhaps under some extraordinary conditions. 
Models that invoke special conditions such
as super-Eddington growth in accretion disks might therefore be
acceptable solutions to the early bottleneck problem.

\subsubsection{Direct Collapse}
The early bottleneck to growth described above arises because 
of the negative feedback from star formation. In principle, the
same is true for the rapid collapse of more massive clouds 
(Loeb \& Rasio 1994; Eisenstein \& Loeb 1995). However,
there is again an intriguing possibility in atomic cooling halos.
If H$_{2}$ and metal cooling were suppressed, atomic hydrogen
cooling could still allow the gas to collapse into the halo with
$T_{\rm vir}\sim 10^4$\,K. But due to the absence of lower temperature
coolants, the collapse would proceed isothermally without any sub-fragmentation,
and therefore without star formation. Recently, the atomic cooling
halo pathway has received considerable attentions, both from the
simulation side (Bromm \& Loeb 2003b; Wise, Turk \& Abel 2008; Regan \& Haehnelt 2009a;
Johnson et al. 2010; Latif, Zaroubi \& Spaans 2011; Shang, Bryan \& Haiman 2010), and with analytical work
(Begelman, Volonteri \& Rees 2006; Lodato \& Natarajan 2006, 2007; Spaans \& Silk 2006).
The key question is whether the gas can indeed remain free of H$_{2}$ molecules
(Dijkstra et al. 2008; Ahn et al. 2009),
and of metals (Johnson, Greif \& Bromm 2008; Omukai, Schneider \& Haiman 2008)
Again, it is important to remember that such a mechanism, where already more
massive seed BHs with $\gtrsim 10^4 M_{\odot}$ form via direct collapse
of a primordial gas cloud, needs to successfully operate only in a few, rare
cases. Indeed, if {\it every} atomic cooling halo were to produce
a massive seed BH in its center at $z\gtrsim 10$, we would exceed
the locally measured total BH mass density (e.g., Yu \& Tremaine 2002).
Fragmentation may also be suppressed by the strong turbulence in 
inflows with high Mach number, where
gas temperatures are significantly {\it below} the virial temperature (Begelman
\& Shlosman 2009). This scenario still needs to be tested, however, with
realistic simulations.
Recently, a qualitatively different variant of massive seed BH formation during direct collapse has been suggested (Mayer et al. 2010). In this model,
two very massive ($\sim 10^{13} M_{\odot}$) halos merge at high redshifts, 
triggering massive inflows into the center of the ensuing potential
well on such a rapid timescale that negative feedback from star formation
has no opportunity to interfere with the BH assembly process.
It is not entirely clear, however, whether such a set-up will occur in
a realistic cosmological setting.

\subsubsection{Other Models}
 Overall, there appears to remain a large uncertainty in these models.
The Pop~III seed model requires a number of optimistic assumptions on
the efficiency of gas accretion and multiple BH mergers, 
whereas the rapid collapse model critically relies on the
assumption that a massive BH does indeed form in a hot, dense gas cloud.
Alternative models for SMBH formation have also been proposed recently.
Primordial stars powered by dark matter annihilation 
(Spolyar et al. 2008; Iocco et al. 2008; Umeda et al. 2009) are suggested to 
have long lifetimes, because they do not consume hydrogen
by nuclear burning. If such objects continued to accrete
the surrounding gas, they could grow to become more massive than $10^{5} M_{\odot}$.
Such very massive ``dark stars'' can be as luminous as $\sim 10^{10} L_{\odot}$,
in principle detectable with {\it JWST} (Freese et al. 2010), and they
can also collapse to massive BHs at their death.

\subsection{SMBH-First Galaxy Coevolution}
It is well-known that in the local Universe, there is a tight 
correlation between the bulge properties of a galaxy and the mass
of its central BH (Gebhardt et al. 2000; Ferrarese \& Merritt 2000). 
Whether or not the same relationship holds in the young Universe 
is an intriguing question. Volonteri \& Natarajan (2009) 
argue that a similar relationship
can be quickly established, and that it would be mainly driven by accretion onto BHs 
after major mergers of the host galaxies.
Coevolution of the first galaxies and early BHs
might be a key in shaping the high-redshift galaxies,
as has been advocated for somewhat lower-redshift galaxies (Di Matteo, Springel \& Hernquist 2005). 
The detailed study of the star-formation history 
of $z>6$ galaxies might provide clues as
to whether star formation was episodic, both within themselves and in their
progenitor systems (e.g., Labbe et al. 2010).

\section{{\it James Webb Space Telescope} SIGNATURE}

The upcoming {\it JWST}, together with the next-generation of 30-40m extremely
large ground-based telescopes, will revolutionize our picture of the
high-redshift Universe. Among the main {\it JWST} science goals is the
detection of light from the first galaxies, and more generally
to elucidate early structure formation at the end of the cosmic dark ages
(Gardner et al. 2006). The key predictions concern the expected flux and
number densities of the first galaxies, enabling us to assess their detectability
with the instruments aboard the {\it JWST} (e.g., Salvaterra, Ferrara \& Dayal
2011). In carrying out these predictions,
a number of challenges still need to be overcome prior to its projected launch
in $\sim 2015$ (see the contributions in Whalen, Bromm \& Yoshida 2010).
We begin by briefly summarizing the {\it JWST} capabilities. A more detailed
discussion is made by Gardner et al. (2006) and Stiavelli (2009).

\subsection{{\it JWST} Instruments and Sensitivities}

The observatory will carry out deep field imaging
with the Near-Infrared Camera
(NIRCam) and the Mid-Infrared Instrument (MIRI), as well as medium-resolution
spectroscopy with the Near-Infrared Spectrograph (NIRSpec) and MIRI.
NIRCam will have a field of view of
$2.2'\times 4.4'$, and an angular resolution of $\sim
0.03''-0.06''$ in the range of observed wavelengths $\lambda_{\rm obs} =
0.6-5 \mu$m.  
The multi-object spectrograph NIRSpec will carry out
medium resolution ($R \sim 100 - 3000$) spectroscopy of up to $\sim
100$ objects simultaneously within a field of view of $3.4'\times
3.4'$, where 
$R \equiv
\lambda_{\rm obs}/\Delta\lambda_{\rm obs}$ is the spectral resolution.
NIRSpec will operate in the same wavelength range as 
NIRCam but at lower angular resolution ($\sim 0.1''$). Finally, 
MIRI will complement NIRCam and NIRSpec by providing
imaging, low and medium resolution spectroscopy within the range of
observed wavelengths $\lambda_{\rm obs} = 5-28.8 \mu$m and fields of
view and angular resolutions of, respectively, $\sim 2'\times 2'$ and
$\sim 0.1''-0.6''$.

In quoting sensitivities, or flux limits $f_{\rm lim}$, for the {\it JWST}
instruments, a signal-to-noise ratio of $S/N=10$ and exposure times of
$t_{\rm exp} =10^4$\,s are often assumed. These baseline sensitivities
are summarized in table~10 by Gardner et al. (2006).
Ultra-deep exposures with {\it JWST}
may extend to $t_{\rm exp} =10^6$\,s, which is comparable to the 
HUDF observations, with flux limits being rescaled according to:
$f_{\rm lim}\propto 1/\sqrt{t_{\rm exp}}$.
Panagia (2005) contains a useful
graphical representation of the {\it JWST} sensitivities, nicely emphasizing
the jump in going from the near-IR to the mid-IR. Approximate numbers, for the
deep exposures, are
$f_{\rm lim}\sim 1$\,nJy for NIRCam, and 10 times higher for the MIRI imager; 
spectroscopic limits are typically two orders of magnitude higher than the imaging
ones. 
It is customary to also work with the AB magnitude system (Oke 1974; Oke \& Gunn 1983).
Specific fluxes, $f_\nu$, can then be expressed as

\begin{equation}
m_{\rm AB} = -2.5 \log_{\rm 10} \left( \frac{f_\nu}{\rm nJy} \right) + 31.4.
\end{equation}

\noindent
Even for
exposure times as long as $10^6$\,s, {\it JWST} will not have
sufficient sensitivity to detect sources with stellar masses below
$\sim 10^5-10^6 M_{\odot}$. In particular, {\it JWST}
will not be able to directly detect individual Pop~III stars at high redshifts
(Bromm, Kudritzki \& Loeb 2001). Therefore, starbursts in the first galaxies are
the primary targets for {\it JWST}. As was already recognized
by Partridge \& Peebles (1967), the first galaxies were likely brightest in
the recombination lines of hydrogen and helium (Schaerer 2002, 2003; Johnson
et al. 2009; Pawlik, Milosavljevic \& Bromm 2011),
in particular the
Lyman-$\alpha$, H$\alpha$ and
He\,II~1640\,\AA 
nebular emission lines (see {\bf Figure~11}).

The flux from the redshifted He\,II~1640\,\AA line ($\lambda_{\rm
em} = 1640$\,\AA), as well as the flux from the redshifted Ly$\alpha$
line ($\lambda_{\rm em} = 1216$\,\AA), would
be detected by {\it JWST} with NIRSpec at a spectral resolution
of $R \sim 1000$, whereas the redshifted H$\alpha$ line ($\lambda_{\rm
em} = 6563$\,\AA) would be detected with MIRI at a spectral
resolution $R \sim 3000$.  
Finally, the redshifted (soft)
UV continuum, at $\lambda_{\rm em} = 1500$\,\AA, would be detected using 
NIRCam.

\subsection{Observing High-redshift Sources}
It is convenient to review the basic relations that relate observed to
intrinsic quantities, as employed in observational cosmology (see also Loeb 2010).

We begin by translating 
intrinsic line and UV continuum luminosities into observed fluxes.
The specific flux from a spatially unresolved object emitted in a
spectrally unresolved line with rest-frame wavelength $\lambda_{\rm em}$
and intrinsic line luminosity $L_{\rm em}$ is given by
Oh (1999) and Johnson et al. (2009):

\begin{equation}
f(\lambda_{\rm obs}) = \frac{L_{\rm em}}{4 \pi d_{\rm L}^2(z)}\frac{1}{\Delta \nu_{\rm obs}}\mbox{\ ,}
\end{equation}
where $\Delta \nu_{\rm obs}=c/(\lambda_{\rm obs}R)$, and
$\lambda_{\rm obs}
= (1+z)\lambda_{\rm em}$. A convenient approximation for the luminosity distance is: $d_{\rm L} \sim
100 [(1+z) / 10] $\,Gpc. For typical parameters, one then has:

\begin{equation}
f(\lambda_{\rm obs}) \simeq
 3 {\rm\,nJy\,} \left(\frac{L_{\rm em}}{10^{40} {\rm erg\,s}^{-1} }\right)\left( \frac{\lambda_{\rm em}}{1216 {\rm \,\AA}}\right) \left(\frac{R}{1000}\right)\left(\frac{1+z}{11}\right)^{-1} \mbox{\ .}\nonumber
\end{equation}

Let us now discuss whether the lines, expected to be emitted by the first 
galaxies, are indeed spatially and spectrally unresolved.
The assumption of spectrally unresolved lines is excellent
for both H$\alpha$ and He\,II~1640\,\AA, whose line widths $\Delta \lambda /
\lambda < 10^{-4} (T/10^4 {\rm K})^{1/2}$ are set by thermal Doppler broadening at
temperature $T < 10^4$\,K (Oh 1999).  
At redshifts $z \gtrsim 10$ a transverse physical scale
$\Delta l$ corresponds to an observed angle $\Delta \theta = \Delta l
/ d_{\rm A} \sim 0.1'' (\Delta l / 0.5 {\rm kpc}) [(1+z)/10]$, where
$d_{\rm A} = (1+z)^{-2} d_{\rm L}$ is the angular diameter
distance. If the recombination lines originate in the 
ionized nebulae in the central regions of the first galaxies at
$r < 0.1 r_{\rm vir}$,
the assumption that the emitting regions are spatially unresolved is also
good for both the H$\alpha$ and the He\,II~1640\,\AA lines, and it applies equally
well to the UV continuum. Here, we use a virial radius of $r_{\rm vir}\sim 1$\,kpc
to describe the overall size of the first galaxies, which is typical for the systems discussed
in Section~4.
In contrast, the Lyman-$\alpha$ line undergoes resonant
scattering (Harrington 1973; Neufeld 1990), and hence will
originate from within a spatially extended
region with typical angular size $\Delta \theta \sim 15''$
(Loeb \& Rybicki 1999), and be heavily damped due to absorption by
intergalactic neutral hydrogen (Santos 2004; but see
Dijkstra \& Wyithe 2010). Indeed, Lyman-$\alpha$ radiation from galaxies
at redshifts $z \gtrsim 10 $ may be severely attenuated
because the bulk of the Universe was likely still substantially neutral at
these redshifts.

A complementary way to quantify the strength of an observed line
uses (redshifted) equivalent widths, which can easily be translated
into the corresponding rest-frame values (e.g., Johnson et al. 2009):
$W_0=f_{\rm line}/f_{\lambda}$, where we have used the intrinsic
line and neigboring (specific) UV continuum fluxes. Predicted equivalent
widths for the first galaxies can reach $W_0\gtrsim 100$\,\AA for He\,II~1640\,\AA,
and $W_0\gtrsim 100$\,\AA for the hydrogen lines (Johnson et al. 2009).

\subsection{Modelling Star Formation in the First Galaxies}

Making predictions for the luminosities and colors of the first galaxies
sensitively depends on what one assumes for the stellar populations
and star formation model (e.g., Schaerer 2002, 2003; Johnson et al. 2009;
Raiter, Schaerer \& Fosbury 2010; Pawlik, Milosavljevic \& Bromm 2011; Salvaterra, Ferrara \& Dayal 2011).
One possibility is that
stars form in a single
instantaneous burst with total stellar mass

\begin{equation}
M_\star \sim 10^5 M_{\odot} \left(\frac{f_{\star}}{ 0.1}\right) \left(\frac{f_{\rm cool}}{0.01}\right) \left(\frac{M_{\rm vir}}{ 10^8 M_{\odot}}\right), 
\label{Eq:Starburst}
\end{equation}
where $f_{\rm cool}$ is a conversion factor that determines the amount of
gas mass available for starbursts inside halos 
with virial masses $M_{\rm vir}$, and $f_\star$ is the star-formation efficiency,
i.e., the fraction of the available gas mass that is turned into
stars. The parameters are normalized to what we have learned from simulating
the formation of atomic cooling halos (see Section~4). Specifically,
the choice of $f_{\rm cool} = 0.01$ reflects the rapid accretion
($t_{\rm acc} < 10$\,Myr) of large
gas masses ($M_{\rm gas} > 10^6 M_{\odot}$) into the central regions, as seen
in the simulations.
The star formation efficiency may be quite high in a burst mode, 
$f_\star = 0.1$, where accretion times are comparable
to the typical lifetimes ($\sim 10$ Myr) of
massive stars. Star formation may then
not be affected by strong feedback capable of
halting the collapse of the accreting gas.
Another possibility is that stars form continuously. Atomic cooling halos,  
with their masses of $\sim 10^8 M_{\odot}$, may have potential wells that
are still too shallow to enable continuous star formation
despite the disruptive effects of stellar feedback (see Section~4).
Galaxies with total (virial) masses of $\gtrsim 10^9 M_{\odot}$, however,
may have been able to sustain such a near-continuous mode (Wise \& Cen 2009).
One can approximately include the effect
of stellar feedback by employing a lower
efficiency, $f_\star = 0.01$, than appropriate for a starburst. The implied
star formation rates $\dot{M}_\star(z) \sim 0.1 M_{\odot}$\,yr$^{-1}$ are
consistent with those found in recent low-mass galaxy
formation simulations (Wise \& Cen 2009;
Razoumov \& Sommer-Larsen 2010).

The luminosities of the first galaxies
critically depend on the metallicities, ages, and IMF of their stellar
populations. Some of the lowest-mass galaxies may still contain
zero-metallicity gas. The resulting stars may form with a
top-heavy IMF, biased
towards high mass ($M_\star \sim 100 M_{\odot}$) stars, as is
expected to be the case for the first, metal-free 
generation of stars which form via molecular hydrogen cooling
(Bromm et al. 2009).
The IMF of metal-free stars is, however, still subject to large
theoretical uncertainties.  Stars forming out of gas with elevated
electron fractions, such as produced behind structure formation or
SN shocks, or as present in ionized regions, could have
characteristic masses substantially less than $< 100
M_{\odot}$ (see Section 4). 
The assumption of metal-free star formation will be violated if
previous episodes of star formation, for instance inside the
progenitors of the assembling galaxy, enriched the gas with
metals. Even a modest enrichment to critical metallicities 
as low as
$Z_{\rm crit} < 10^{-6} - 10^{-3.5} Z_{\odot}$ may imply the transition from a
top-heavy IMF to a normal IMF (Bromm et al. 2001; Santoro \& Shull 2006;
Schneider et al. 2006; Smith \& Sigurdsson 2007). 
Note that even a few SN explosions
may already be sufficient to enrich low-mass
($\sim 10^8 M_{\odot}$) galaxies to metallicities $Z > Z_{\rm crit}$
(Wise \& Abel 2008; 
Karlsson, Johnson \& Bromm 2008; Greif et al. 2010; Maio et al.
2011).

The luminosity in the He\,II~1640\,\AA line strongly depends
on both the IMF and stellar metallicity, 
and also on the age of the galaxy, i.e., the time since the last 
major star-formation episode.
At fixed IMF, a change from low to zero metallicity implies an increase in the
He\,II~1640\,\AA line luminosity by about three orders of magnitude 
for the first few million years after the starburst.
This reflects the exceptionally hot atmospheres of zero-metallicity stars that
render them into strong emitters of He\,II ionizing radiation 
(Tumlinson \& Shull 2000; Bromm, Kudritzki \& Loeb 2001;
Schaerer 2003).  
For a top-heavy IMF, as advocated for primordial or
very low-metallicity stars, 
the line luminosity is increased by another order of magnitude (see 
{\bf Figure~\ref{RECOMB}}).
The large differences in luminosities
offer the prospect of distinguishing observationally between stellar
populations consisting of metal-free or metal-enriched stars, and of
constraining their IMFs (Tumlinson \& Shull 2000;
Bromm, Kudritzki \& Loeb 2001; Oh 2001; Johnson et al. 2009).
{\it JWST} has
the potential to constrain the properties of starbursts in galaxies
with halo masses as low as $\sim 10^9 M_{\odot}$, based on the
simultaneous detection/non-detection of the H$\alpha$ and He\,II~1640\,\AA
lines (Pawlik, Milosavljevic \& Bromm 2011).
Indeed, only zero-metallicity starbursts with a top-heavy IMF
can be detected in both H$\alpha$ and He\,II~1640\,\AA, assuming exposure times
$\lesssim 10^6$\,s. Whether Lyman-$\alpha$ can be detected as well will depend
on the attenuation due to
resonant scattering in the neutral IGM.
Because of the greater sensitivity
of NIRSpec compared to MIRI, Lyman-$\alpha$ line emission
is potentially easier to detect than H$\alpha$, and it 
hence remains a very powerful probe of
galaxy formation at redshifts $z \gtrsim 10$, despite the large
uncertainties caused by its resonant nature.

\subsection{Source Number Counts}

The second key prediction concerns the number density of the first
galaxies that {\it JWST} may observe. We can estimate the number
of galaxies detectable with {\it JWST}, per unit solid angle, above redshift
$z$ as follows (e.g., Pawlik, Milosavljevic \& Bromm 2011):

\begin{equation}
\frac{dN}{d \Omega}(>z) = \int_z^\infty dz'\ \frac{dV}{dz'd\Omega}
\frac{\tau_{\rm sb}}{t_{\rm H}(z')}\int_{M_{\rm min} (z')}^\infty dM\ n(M, z'),
\end{equation}

\noindent
where $t_{\rm H}(z)$ is the age of the Universe at $z$, and

\begin{displaymath}
\frac{dV}{dz d\Omega}=\frac{c d_{\rm L}^2}{1+z}\left|\frac{dt}{dz}\right|
\end{displaymath}

\noindent
the comoving volume element per unit solid angle and redshift.
Here $\left|dt/dz\right|^{-1}\simeq (1+z) H_0 \Omega_{\rm m}^{1/2}
(1+z)^{3/2}$, valid for high redshifts.
$n(M,z)$ is the comoving
number density of galaxy host halos with mass $M$ at redshift $z$, which can
be derived from large cosmological simulations, or calculated with
approximate analytical techniques. The latter approach often relies on variants of the Press-Schechter formalism
(Press \& Schechter 1974; for a recent review, see Zentner 2007).
$M_{\rm min}(z)$ is the lowest (total or virial) halo mass capable of
hosting a starburst that can be detected by the {\it JWST}. It depends
on the stellar properties (metallicity and IMF), and on whether observations
are made in, e.g., the H$\alpha$ line, the He\,II~1640\,\AA line, or in
the soft continuum. Typical values are $M_{\rm min}\sim
10^8 - 10^9 M_{\odot}$ for $z\simeq 10 - 15$ (Pawlik, Milosavljevic \& Bromm 2011).
Finally, $\tau_{\rm sb}$ gives the duration of the starburst, which may
vary from $\sim 3$\,Myr for top-heavy Pop~III stars, to ten times larger
values for stars with normal IMF. In each case, this timescale measures
the approximate time after which negative stellar feedback terminates
the starburst. In {\bf Figure~\ref{COUNTS}}, we show results from a
Press-Schechter based calculation (Pawlik, Milosavljevic \& Bromm 2011), demonstrating that
{\it JWST} may detect a few tens (for $Z>0$ and normal IMF) up to
a thousand (for Pop~III with a top-heavy IMF) starbursts from $z>10$
in its field-of-view of $\sim 10$~arcmin$^2$. This estimate is consistent
with previous studies for similar assumptions about the conversion
between halo and stellar mass (e.g., Haiman \& Loeb 1997, 1998; Oh 1999;
Trenti \& Stiavelli 2008). Current calculations, however,  still suffer from
a number of uncertainties, such as whether Case~B recombination theory
is appropriate in the first galaxies (Schaerer 2003; Raiter, Schaerer
\& Fosbury 2010), the role of dust extinction (Trenti \& Stiavelli 2006),
the feedback-regulated star formation efficiency, and the escape fraction
of ionizing radiation (Gnedin, Kravtsov \& Chen 2008; Wise \& Cen 2009;
Johnson et al. 2009; Razoumov \& Sommer-Larsen 2010; Yajima, Choi \&
Nagamine 2011). 

\section{STELLAR ARCHAEOLOGY}

Stellar Archaeology is the endeavor to constrain the properties of the first stars 
by scrutinizing the chemical abundance patterns in the most metal-poor, and 
therefore presumably oldest, stars in the Milky Way and nearby galaxies
within the Local Group (Beers \& Christlieb 2005; Frebel 2010). 
Such a near-field cosmological approach nicely complements the
traditional far-field cosmology based on high-redshift observations 
(Freeman \& Bland-Hawthorn 2002). The first galaxies may have left behind 
a number of local fossils as well. {\it (i)} Some of the numerous dwarf 
galaxies in the Local Group may constitute the survivors of the first galaxies.
In this regard, the ultra-faint dwarf (UFD) galaxies, recently discovered
in the SDSS, are
of particular promise. {\it (ii)} The first galaxies likely were the
formation sites for the first low-mass Pop~II stars (e.g., Tumlinson 2010). These eventually
found their way into the halo, and possibly bulge, of our Galaxy through
its complex hierarchical assembly process. {\it (iii)} Finally, a subset of
the first galaxies may have provided the birth places for old, metal-poor
globular clusters (GCs), which again might have been incorporated into
our MW (Bromm \& Clarke 2002; Kravtsov \& Gnedin 2005; Brodie
\& Strader 2006; Boley et al. 2009).
We focus on the first issue, as it is of most direct relevance for this 
review.

\subsection{Ultrafaint Dwarf Galaxies}
The newly discovered UFD galaxies are the intrinsically least luminous 
members
($L_{\rm tot} \lesssim 10^{5}\,L_{\odot}$) of the Local Group (Kirby et al. 2008; Martin, de Jong \& Rix 2008).
Due to their simple assembly history, they can be regarded as the
closest local relatives to the first galaxies. They are
believed to have had only one or few early star formation events, but
have been quiescent ever since (Tolstoy, Hill \& Tosi 2009). Hence, they
should reflect the
signatures of the earliest stages of chemical enrichment in their
population of low-mass stars.
As opposed to the MW halo, which was assembled
through numerous merger and accretion events, the lowest luminosity dwarfs
provide us with a much cleaner fossil record of early star and galaxy formation.
With their small
number of stars (of order a few hundred), the UFDs 
may allow us to carry out a virtually
complete census of their stellar content (Simon et al. 2011).
Medium-resolution spectroscopic studies have shown that
all of the UFDs have large [Fe/H] spreads of $\sim1$\,dex 
or more (Kirby et al. 2008; Norris et al. 2010), reaching below
$\mbox{[Fe/H]}=-3.0$. Moreover, some of them have average
metallicities as low as $<\mbox{[Fe/H]}>\sim-2.6$, which is lower than
the values found in the most metal-poor GCs.
The abundances of dwarf
galaxy stars closely resemble those found in similarly metal-poor
Galactic halo stars. Overall, this suggests that chemical evolution proceeded
very similarly at
the early times which are probed with the
most metal-poor, and thus presumably the oldest, stars in a given
system (Frebel \& Bromm 2011). The same chemical behavior has also been found in Sculptor, a more
luminous, classical dwarf spheroidal (dSph) galaxy, at
$\mbox{[Fe/H]}\sim-3.8$ (Frebel et al. 2010). However, at higher metallicity
($\mbox{[Fe/H]}>\sim-2.5$), the Sculptor stellar
([$\alpha$/Fe]-) abundances deviate with respect to the behavior of
Galactic halo stars (Geisler et al. 2005), indicating a different
evolutionary timescale and multiple star-formation events (Tolstoy et al. 2004).

\subsection{Theoretical Models}
There is widespread consensus that the UFDs may provide us with the 
{\it Rosetta Stone} for galaxy formation, given their relative simplicity.
It is therefore very tempting to theoretically model their formation
process. When did they form, and how do they fit into the hierarchical
$\Lambda$CDM cosmology? What kind of star formation history did they
experience, and, related to this, how many SNe did contribute to their
complement of metals? This field is still very young, and it is likely
that progress over the next few years will be rapid. Here, we only
provide a few comments to illustrate the flavor of the developing
argument.
\subsubsection{Formation Site}
Currently, two main ideas for the origin of the UFDs are discussed in 
the literature. One class of models invokes H$_{2}$-cooling minihalos
(Bovill \& Ricotti 2009, 2011; Salvadori \& Ferrara 2009).
The models couple a representation of the evolving dark matter distribution,
either from cosmological simulations or from Press-Schechter type techniques,
with a recipe for star formation and feedback, and can successfully explain
the broad observational properties of the UFD population (see {\bf Figure~\ref{UFD}}).
The suggested antecedents of the UFDs would then have been minihalos with masses
$M\simeq 10^7 - 10^8 M_{\odot}$, close to the threshold where
atomic cooling sets in. A challenge for these models comes from the
highly-resolved, ab initio simulations discussed in Section~4. The 
underlying question again is where second-generation star formation can occur,
already in minihalos or only in the next stage of hierarchical assembly,
the atomic cooling halos (see the discussion in Section~2).
Within the minihalo scenario, the same system would have to first lead
to the explosion of Pop\,III SNe, subsequently reassemble the enriched
gas inside their shallow potential well despite strong negative feedback effects, and finally trigger a second
generation of star formation. The strength of the negative feedback crucially depends 
on the Pop~III IMF; the more top-heavy it is, the longet the delay time between
first and second generation star formation. For the minihalo model as UFD
progenitors to work, one has to assume that the first stars typically were not
too massive.

The above challenge provides the motivation for the competing model to explain the
origin of UFDs (Maccio et al. 2010; Frebel \& Bromm 2011).
In the atomic cooling halo pathway,
the sites for first and second-generation star formation are
decoupled (see {\bf Figure~\ref{ASSEMBLY}}),
thus alleviating the problem of admitting local Pop\,III pre-enrichment.

\subsubsection{Enrichment Mode}

An important clue to the true nature of the UFD formation site could
come from a knowledge of the chemical enrichment mode. Did enrichment
in the UFD progenitors occur in one intial burst, to be completely
shut-off subsequently, or continuously, spread out over an extended
star formation and SN history? The first possibility has been termed
``one-shot'' chemical enrichment by Frebel \& Bromm (2011).
The answer to this question would provide us with important clues
about the strength of the feedback in the first galaxies. If this
feedback was sufficiently violent to disrupt the first galaxy already
after its initial starburst, blowing all remaining gas into the
general IGM, ``one-shot'' conditions would be realized. The simulations
have not yet answered this question with any degree of certainty, but
one can look for the chemical signature of such burst-like enrichment
in the stellar content of the UFDs (Frebel \& Bromm 2011). 
Their surviving Pop\,II stars would then
preserve the yields from the initial Pop\,III SNe that had occurred in
the progenitor minihaloes without any subsequent enrichment from
events that operated on timescales longer than the short dynamical time
that governs the formation of the starburst,
such as type\,Ia SNe or AGB winds. Specifically, one would expect
high [$\alpha$/Fe] values for {\it all} stars in the UFD, and low
n-capture abundances due to the absence of any s-process contribution
from AGB stars. 

An important caveat is that a subset of those Pop~II stars might have
experienced post-processing of their surface abundance, e.g., via 
mass transfer from a binary companion or dredge-up events during later
stages of stellar evolution. A possible strategy to circumvent this
problem is to realize that almost all stars form in clusters. A properly
defined multi-dimensional abundance space could thus uniquely identify the
primordial signature through this clustering effect (Bland-Hawthorn et al. 2010).

\subsubsection{Lessons Learned}
Currently, the lowest luminosity dwarfs are consistent
with the one-shot criteria, but the data is still very sparse,
and the case therefore remains inconclusive.
The hope is that high-resolution spectroscopy of more UFD stars will
soon become available.
The abundance ratios in most individual stars reflect an enrichment
history that is dominated by core-collapse SNe, even in the higher
metallicity regime ($\mbox{[Fe/H]}\sim-2.0$). The latter is dominated by
SN\,Ia enrichment in the more luminous classical dSphs. 
The observed spread in Fe and other elements may suggest that
mixing in the UFD progenitors was not very efficient, at least
on scales of $\gtrsim 10$\,pc, whereas mixing on smaller scales may
have been almost complete, if the simulations discussed in Section~4 are correct.
The suggested signature from clustered star formation in the first galaxies
may again help to constrain the mixing efficieny on different length scales
(Bland-Hawthorn et al. 2010).
Without inhomogeneous mixing, all stars
should have nearly identical abundances, similar to what is found in
globular cluster. We can thus tentatively infer that GCs
must have
formed in more massive haloes where turbulent mixing would have been
much more efficient.

As additional abundances of individual dwarf galaxy stars
become available, abundance gradient studies of the UFD galaxies
should shed further light on the mixing efficiency. Stronger
gravitational fields in the center of a system would drive more
turbulence that in turn would induce mixing. Because the UFDs are ideal
testbeds for various feedback processes, it will also be interesting
to study the carbon abundances in these systems. Carbon, as well as
oxygen, may have been a key cooling agent inside the first galaxies
(Bromm \& Loeb 2003a). Although one extremely carbon-rich star (with
$\mbox{[Fe/H]}\sim-3.5$) has recently been found in Segue\,1
(Norris et al. 2010a), low
stellar C abundances, if ever found, would greatly weaken the theory of
fine-structure line cooling for driving the transition to low-mass
star formation.

\section{OUTLOOK}

The most crucial immediate challenge, for both observers and theorists,
is to close the gap between the mass scale accessible to ab initio
simulations (virial masses of $\sim 10^8 M_{\odot}$), and to cutting-edge observations
(inferred total masses of $\sim 10^{10} M_{\odot}$).
We have encountered this
fundamental problem repeatedly in our preceeding discussion.
A second key need is to derive better predictions for the number counts of 
the first galaxies, and to devise robust multi-color and spectroscopic
criteria to disentange the likely mix of Pop~III and Pop~II stars, possibly
together with an AGN component, encountered in the first galaxies.
The appearance of the first galaxies in sub-millimeter to radio bands needs 
to be explored theoretically. In particular, atomic and molecular lines 
such as CII and CO lines may be promising in detecting and characterizing 
the first galaxies (Walter \& Carilli 2007; Obreschkow et al. 2009).  
Finally, to fully harness the tremendous potential of stellar archaeology in local
dwarf galaxies, a much increased sample of 
high-quality elemental abundances is needed.

The study of the first galaxies enters an exciting period, where
advances in supercomputer technology enable ever more realistic
ab initio simulations within a realistic cosmological context.
This is matched by equally exciting prospects on the observational side,
where next-generation facilities -- such as {\it JWST}, the planned 30-40m
extremely large telescopes on the ground, ALMA, and the SKA -- will finally
open up the high-redshift frontier. It is very likely that if another 
review on the first galaxies is written a decade from now, our
understanding of the subject will have completely changed. This again
reflects the special stage this field is in, where we are just
at the threshold of a golden age of discovery.

\section*{ACKNOWLEDGMENTS}
\noindent
The authors thank Andrew Benson, Andrea Ferrara, Zoltan Haiman,
Adam Lidz, Masami Ouchi, and John Wise for carefully reading earlier 
drafts of the present review and also for providing comments.
V.B. is supported by the National Science Foundation grant AST-1009928,
and by NASA through Astrophysics Theory and Fundamental Physics Program
grants NNX08-AL43G and NNX09-AJ33G. 
N.Y. acknowledges support from
the Grants-in-Aid for Young Scientists (S) 20674003
by the Japan Society for the Promotion of Science and 
from World Premier International Research Center Initiative (WPI Initiative), 
MEXT, Japan.

\begin{figure}[ht]
\begin{center}
\includegraphics[trim=1in 1in 1in 1in, clip ,width = 5 in] {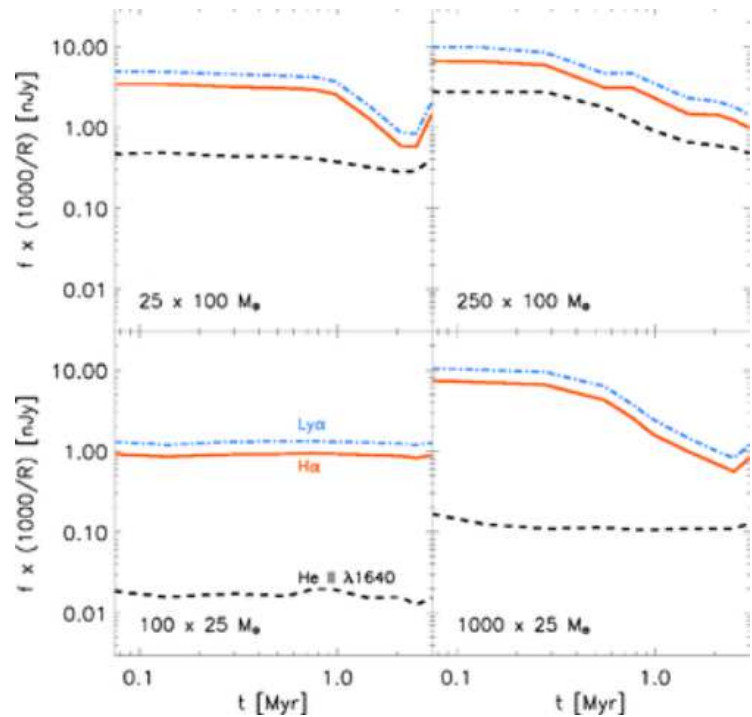}
\vspace{-1.5 in}
\caption{Emission line fluxes in the first galaxies. Shown are predictions for observable
recombination line fluxes as a function of time. The source is an
atomic cooling halo at $z\simeq 12.5$. The lines are: Ly$\alpha$ (dot-dashed
blue), H$\alpha$ (solid red) and He\,II~1640\,\AA (dashed black).
The fluxes are normalized to a spectral resolution of $R=1000$.
The Ly$\alpha$ flux is an upper limit, due
to the possibly severe attenuation by the surrounding, still largely
neutral, IGM.
Adopted from Johnson et al. (2009).}
\label{FLUX}
\end{center}
\end{figure}

\begin{figure}[ht]
\begin{center}
\includegraphics[trim=1in 1in 1in 1in, clip ,width = 5 in] {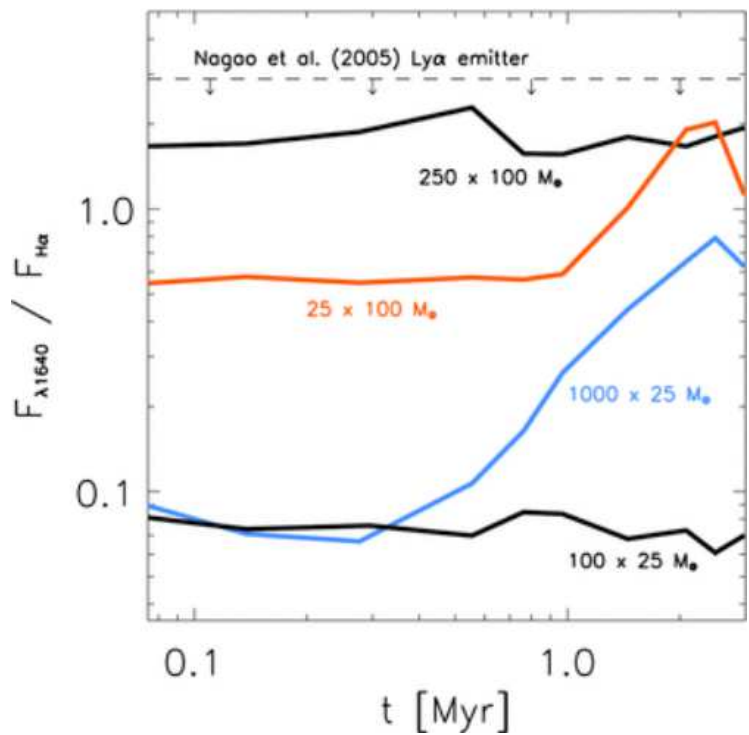}
\vspace{-1.5 in}
\caption{IMF diagnostics in the first galaxies. Shown is the flux ratio
in the He\,II~1640\,\AA to H$\alpha$ recombination line as a function
of time. The calculation assumes a central cluster of Pop~III stars, all
either with a mass of $25 M_{\odot}$ or $100 M_{\odot}$ for simplicity.
The more massive Pop~III stars lead to a ratio that is an order of
magnitude larger, thus enabling to diagnose the nature of the stellar
population. The dashed horizontal line corresponds to the upper limit
for the strong Lyman-$\alpha$ emitter SDF~J132440.6+273607 at $z\simeq 6.3$
(Nagao et al. 2005). Evidently, this limit does not yet allow to distinguish
between different populations. Adopted from Johnson et al. (2009).}
\label{RECOMB}
\end{center}
\end{figure}

\begin{figure}[ht]
\begin{center}
\includegraphics[width = 5 in] {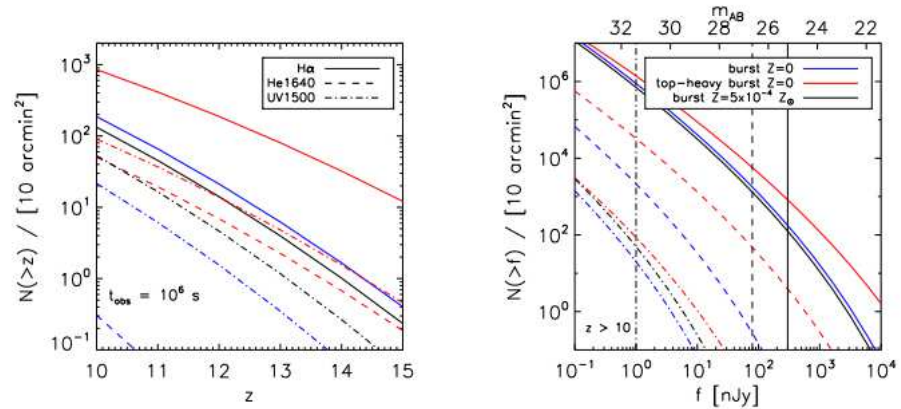}
\caption{{\it JWST} number counts of the first galaxies.
The calculations assume $t_{\rm exp}=10^6$\,s and $S/N=10$.
{\it Left panel:} Number of galaxies $N(>z)$ with redshifts $z>10$ hosting
a starburst observable through the detection of H$\alpha$ (solid lines),
He\,II~1640\,\AA (dashed lines), or the soft UV continuum (dash-dotted
lines). Colors denote
different choices for stellar metallicity and IMF, as described in the inset
of the righ-hand panel.
{\it Right panel:} Number of galaxies $N(>f)$ above $z>10$ with observed
fluxes $>f$. The vertical lines show the {\it JWST} flux limits for
H$\alpha$ (solid),
He\,II~1640\,\AA (dashed), and the soft UV continuum (dash-dotted).
{\it JWST} may detect a few tens (for $Z>0$ and normal IMF) up to
a thousand (for Pop~III with a top-heavy IMF) starbursts from $z>10$
in its field-of-view of $\sim 10$~arcmin$^2$.
Adopted from Pawlik, Milosavljevic \& Bromm (2011).}
\label{COUNTS}
\end{center}
\end{figure}

\begin{figure}[ht]
\begin{center}
\includegraphics[width = 5 in] {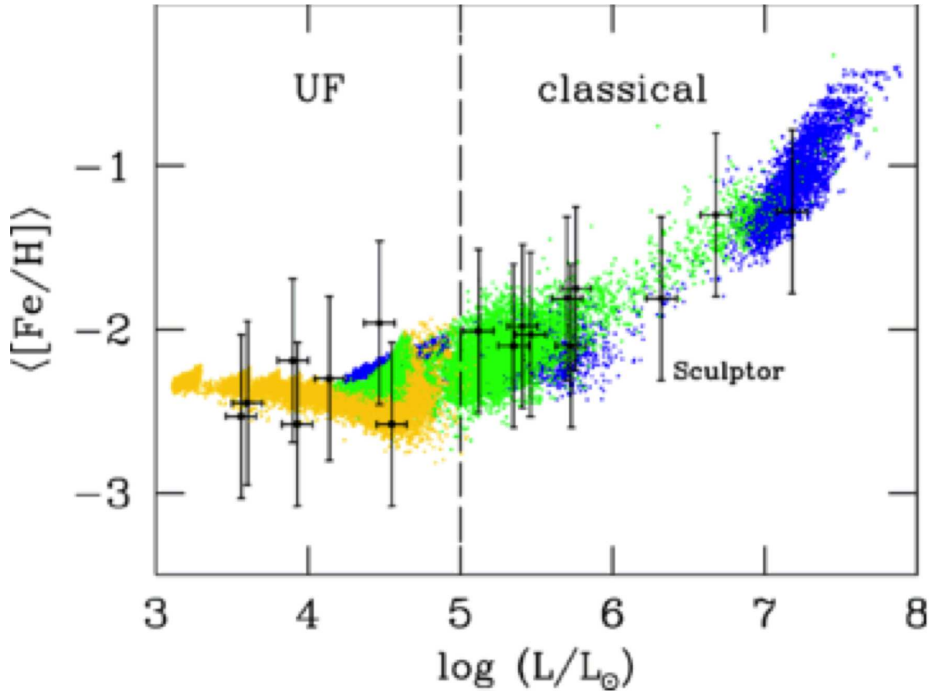}
\vspace{-1.5 in}
\caption{Stellar archaeology with dwarf galaxies.
Shown are average Fe abundances vs. total luminosities for dwarf galaxies
as predicted by a semi-analytical merger tree model. Different colors indicate
the baryon fraction at the time of formation, expressed relative to the cosmic
mean: $f_{\rm b}/\bar{f}_{\rm b} > 0.5$ (blue dots),
$0.1 < f_{\rm b}/\bar{f}_{\rm b} < 0.5$ (green),
and $f_{\rm b}/\bar{f}_{\rm b} < 0.1$ (yellow). The symbols with erroe bars
denote observational data from Kirby et al. (2008).
Within this model, the ultra-faint (UF) dwarf galaxies are the fossils of
minihalos with (virial) masses close to the limit where atomic cooling would set in
($M\simeq 10^7 - 10^8 M_{\odot}$). The classical dwarf spheroidals, such as 
the prototypical Sculptor system, would then be descendants of more massive
dark matter halos.
Adopted from Salvadori \& Ferrara (2009).}
\label{UFD}
\end{center}
\end{figure}
\end{document}